%% file: ms.tex
\documentclass[10pt,preprint2,usenatbib]{aastex}

\newcommand{\mum}{\ifmmode{\rm \mu m}\else{$\mu$m}\fi}

\newcommand{\chisq}{\ifmmode{\chi^{2} }\else{$\chi^2$}\fi}
\newcommand{\rchisq}{\ifmmode{\chi^{2} }\else{$\chi^2_\nu$}\fi}

\usepackage{amssymb}
\usepackage{graphicx}
\usepackage{threeparttable}
\usepackage{multirow}
\usepackage{multicol}
\usepackage{url}
\usepackage{color}

\shorttitle{Evolved Stars in Sextans A \& Leo A}  \shortauthors{O.~C.~Jones et al.}                

\begin{document}

\title{Near-Infrared Stellar Populations in the metal-poor, Dwarf irregular Galaxies Sextans A and Leo A}
\author{Olivia~C.~Jones\altaffilmark{1,6},
        Matthew~T.~Maclay \altaffilmark{1,2},
        Martha~L.~Boyer\altaffilmark{1},  
        Margaret~Meixner\altaffilmark{1,3},
        Iain McDonald\altaffilmark{4},
        Helen Meskhidze\altaffilmark{1,5}
        }
\email{ojones@stsci.edu}
\altaffiltext{1}{Space Telescope Science Institute, 3700 San Martin Drive, Baltimore, MD 21218, USA}
\altaffiltext{2}{Carleton College, 1 North College Street, Northfield, MN 55057, USA}
\altaffiltext{3}{The Johns Hopkins University, Department of Physics and Astronomy, 366 Bloomberg Center, 3400 N. Charles Street, Baltimore, MD 21218, USA}
\altaffiltext{4}{Jodrell Bank Centre for Astrophysics, Alan Turing Building, School of Physics and Astronomy, The University of Manchester, Oxford Road, Manchester, M13 9PL, UK}
\altaffiltext{5}{Elon University, 100 Campus Drive, Elon, NC 27244, USA}
\altaffiltext{6}{UK Astronomy Technology Centre, Royal Observatory, Blackford Hill, Edinburgh, EH9 3HJ}

\begin{abstract}
\noindent We present \emph{JHK}$_{s}$ observations of the metal-poor ([Fe/H] $<$ --1.40)  Dwarf-irregular galaxies, Leo A and Sextans A obtained with the WIYN High-Resolution Infrared Camera at Kitt Peak. Their near-IR stellar populations are characterized by using a combination of colour-magnitude diagrams and by identifying long-period variable stars.
We detected red giant and asymptotic giant branch stars, consistent with membership of the galaxy's intermediate-age populations (2--8 Gyr old). Matching our data to broadband optical and mid-IR photometry we determine luminosities, temperatures and dust-production rates (DPR) for each star. We identify 32 stars in Leo A and 101 stars in Sextans A with a DPR $>10^{-11}$  $M_\odot \,{\rm yr}^{-1}$, confirming that metal-poor stars can form substantial amounts of dust. We also find tentative evidence for oxygen-rich dust formation at low metallicity, contradicting previous models that suggest oxygen-rich dust production is inhibited  in  metal-poor environments. The total rates of dust injection into the interstellar medium of Leo A and Sextans A are (8.2 $\pm$ 1.8) $\times 10^{-9}$  $M_\odot \,{\rm yr}^{-1}$ and (6.2 $\pm$ 0.2) $\times 10^{-7}$ $M_\odot \,{\rm yr}^{-1}$, respectively. The majority of this dust is produced by a few very dusty evolved stars, and does not vary strongly with metallicity. 
\end{abstract}


\section{Introduction}
\label{sec:intro}

Towards their end of their nuclear-burning lives, stars of initial mass $ 1 M_\odot \lesssim M \lesssim 8 {\rm M_\odot} $ evolve along the asymptotic giant branch (AGB). During this short evolutionary phase \citep[$t \sim 0.5-10$ Myr;][]{Vassiliadis1993, Girardi2007}, AGB stars are among the most luminous infrared (IR) sources in a galaxy \citep{Maraston2006, Melbourne2012}, they thermally pulsate and experience significant mass loss of up to 10$^{-4}$ M$_{\odot}$ yr$^{-1}$. Mass-loss from AGB stars increases as the star evolves. This gas and dust chemically enriches the interstellar medium (ISM).

During the thermally pulsating AGB (TP-AGB) phase, the stellar atmosphere can transition from oxygen rich to carbon rich via the dredge-up process from the stellar core \citep{Iben1983}. Oxygen-rich AGB stars have an atmospheric C/O $<$ 1, whilst Carbon stars have C/O $>$ 1. This transition depends on the star's initial mass and metallicity. At solar metallicity, stars with an initial mass of $\sim$2--4 ${\rm M}_{\odot}$ form carbon stars \citep{Marigo2007, Ventura2010}. At lower metallicities there is less initial oxygen available, thus fewer dredge-up events are required to form carbon stars,  resulting in a wider range of progenitor masses \citep{Ferrarotti2001, DiCriscienzo2013}.
This has a significant effect on the stars' IR colours. Near-IR wavelengths, where the star emits the majority of its flux, can be used to investigate a galaxy's intermediate-age population, which can be used to trace their baryonic enrichment of the host galaxy.

Metal-poor ($-2.1 \lesssim [{\rm Fe/H}] \lesssim  -1.1$) dust-producing AGB stars have been observed in the mid-IR by the DUST in the Nearby Galaxies with Spitzer (DUSTiNGS) survey \citep{Boyer2015a, Boyer2015b}. This survey used variability criteria to identify luminous AGB stars in 50 dwarf galaxies within 1.5 Mpc. These galaxies offer a fantastic opportunity to investigate evolved stellar populations over a wide range of environments, star-formation histories (SFH) and metallicity. In this work, we focus on the near-IR population of two galaxies in the DUSTiNGS survey, Sextans A and Leo A. Their properties are listed in Table~\ref{tab:galProp}.

Sextans A (DDO 75) is an extremely metal-poor ($[{\rm Fe/H}] \sim -1.85$ to $-1.40$ ; \citealt{Dolphin2003b}), dwarf irregular galaxy. Located at a distance of 1.45 Mpc \citep{McConnachie2012}, it is a member of the NGC 3109 association \citep{vandenBerghBook}, which may or may not be bound to the Local Group.
Photometric studies of Sextans A reveal that the galaxy contains an old stellar population (which formed 10-14 Gyrs ago), and a younger blue population which formed during the past 2.5 Gyr \citep{DohmPalmer2002}.  
The star formation rate of Sextans A has increased dramatically in the past few Gyr; from a quiescent period between 2.5 and 10 Gyr ago, to a sudden burst in star-formation starting 1--2.5 Gyr ago, with another considerable increase over the past 0.06 Gyr \citep{Dolphin2003b, Weisz2014, Camacho2016}. There has been little chemical enrichment within Sextans A over the last 10 Gyr: 
Sextans A is one of the lowest-metallicity objects in which variable AGB stars have been detected \citep{Boyer2015b, McQuinn2017}. The galaxy also contains a planetary nebula \citep{Magrini2005b} and several short-period Cepheids \citep{Dolphin2003a}.

Leo A (also named Leo III or DDO 69) is an isolated, dwarf irregular galaxy in the Local Group, at a distance of about 790 kpc \citep{Bernard2013, Dolphin2002}. This gas-rich galaxy \citep{Young1996} contains both young and old populations \citep{Cole2007} and is thought to be a dark-matter-dominated ($>$80\%) stellar system \citep{Brown2007}. 
Leo A is metal poor ($[{\rm Fe/H}] \sim -2.1$ to $-1.2$; \citealt{Kirby2017}), and the majority of its stellar population is of a young to intermediate age, forming within the last 5--8 Gyr. The last major episode of star formation occurred around 3 Gyr ago. Ongoing star-formation activity is traced by three H\,{\sc ii} which are found in the galaxy \citep{Strobel1991, Weisz2014}. Whilst the detection of RR Lyr stars by \citet{Dolphin2002} indicates that an ancient stellar population formed $> 11$ Gyr ago, possibly during the era of reionization. 
Furthermore, five variable AGB candidates were identified by \citet{Boyer2015b} and a planetary nebula was discovered by \citet{Magrini2003}.

\begin{table}
  \small
\begin{center}
\caption{Sextans A and Leo A Parameters Adopted in this Work}
\label{tab:galProp}
\begin{tabular}{lcc}
\hline
\hline
Parameter	            &   Sextans A	 &      Leo A		\\
\hline
RA                          & 10:11:00.8     &      09:59:26.5	        	\\
Dec	                    & --04:41:34     &     +30:44:47	        	\\
Distance, $d$ (kpc)	    & 1432 $\pm$ 53  &      789  $\pm$  44      	\\
$(m-M)_0$ (mag)	            & 25.6 $\pm$ 0.1 &     24.51 $\pm$ 0.12     	\\
$[\rm{Fe/H}]$    	    &  --1.85 	     &   --1.4 (2)	    	\\
Half-light radius ($^\prime$)  &  2.47		     &     2.15		       	\\

Stellar Mass (10$^6$ M$_\odot$)   &   44          &       6                      \\
$E(B-V)$ (mag)                  &  0.045 	&     0.021	        	\\
$I$ TRGB (mag)       	        &  21.76        &      20.5                     \\
\hline
\tablenotetext{1}{The fundamental parameters of the Sextans A and Leo A galaxies \\
  are taken from \citet{McConnachie2012} unless otherwise noted.\\ (2) \cite{Kirby2017}.}
\end{tabular}
\end{center}
\end{table}

The goal of this paper is to identify the dusty evolved stellar populations in metal-poor irregular galaxies, Sextans A and Leo A.
This paper is organized as follows.
In Section~\ref{sec:data}, we describe the WHIRC observations and data reduction. 
The photometric results and colour magnitude diagrams are presented in Section~\ref{sec:Results}.
We analyze these results in Section~\ref{sec:Discussion} and compare these findings with other studies.
Section~\ref{sec:conclusion} contains a summary and conclusions.

\section{Observations and Data reduction}
\label{sec:data}

\subsection{Observations}
\label{sec:Obs}

Ground-based observations of Sextans A and Leo A were obtained through the broadband \emph{JHK}$_{s}$ filters using the WIYN High-Resolution Infrared Camera (WHIRC; \citealt{Meixner2010}) mounted on the 3.5m WIYN telescope at the Kitt Peak National Observatory. The central wavelengths and bandwidths widths of the three filters can be found in Table~\ref{tab:filtertable}. The WHIRC camera field of view is 3.4'$\times$ 3.4' with a plate scale of $\approx$ 0.1'' pixel$^{-1}$.  Details of the observations can be found in Tables~\ref{table:obslogSexA} and \ref{table:obslogLeoA}. The data were collected over four observing runs over a period spanning from 2010 May to 2011 April.  Observations were centered around RA = 10\textsuperscript{h}08\textsuperscript{m}29\textsuperscript{s}.5 and Dec $= -0^{\circ}26\textsuperscript{m}45\textsuperscript{s}$
for Sextans A, and RA = 09\textsuperscript{h}59\textsuperscript{m}26.1\textsuperscript{s} and Dec = $+$30$^{\circ}$44\textsuperscript{m}55.5\textsuperscript{s} for Leo A. Dedicated background control fields were also observed.

\begin{table}
\begin{center}
\caption{WHIRC Filter Characteristics}
\label{tab:filtertable}
\begin{tabular}{lcc}
 & & \\  
\hline
\hline
Filter          &   $\lambda_{c}$($\mu$m)   &     $ \Delta\lambda$($\mu$m)      \\
\hline
J               &   1.250                   &     0.1619                       \\
H               &   1.651                   &     0.3009                       \\
K$_{\rm{s}}$      &   2.150                   &     0.3430                        \\
\hline
\end{tabular}
\end{center}
\end{table}

\begin{table*}
\centering
\caption{Observation Log for Sextans A}
\label{table:obslogSexA}
\begin{tabular}{cccccc}
\hline
\hline
Band                    &  Observation Date  &  Interval (days) &   Exp.~Time (s)  &   Seeing FWHM ('')   &   Airmass      \\  
                                                      
\hline                                                
\em{J}                  &   2010-05-02        & 0               &   120                &   1.8           &   1.35 -- 1.45    \\  
\em{J}                  &   2011-01-23        & 266             &   180                &   1             &   1.70 -- 1.95    \\  
\em{J}                  &   2011-01-24        & 267             &   180                &   1.1           &   1.72 -- 2.57    \\  
\em{J}                  &   2011-04-15        & 348             &   180                &   \ldots        &   1.26 -- 1.41    \\  
\em{H}                  &   2011-01-24        & \ldots          &   180                &   1.6           &   1.30 -- 1.43    \\  
\em{K$_{\rm{s}}$}         &   2011-01-23        &  0              &   60                 &   0.7           &   1.25 -- 1.33    \\  \em{K$_{\rm{s}}$}         &   2011-01-23        &  0              &   80                 &   0.7           &   1.48 -- 1.25    \\  \hline
\tablenotetext{1}{Interval is the separation in days between the first observational epoch and latter epochs.}
\end{tabular}
\end{table*}

\begin{table*} 
\caption{Observation Log for Leo A}
\centering
\label{table:obslogLeoA}
\begin{tabular}{cccccc}
\hline
\hline
Band                    &  Observation Date   &  Interval (days) &   Exp.~Time (s)   &   Seeing FWHM ('')           &   Airmass          \\  
\hline                                                
\em{J}                  &   2010-01-22        & 0               &   180              &   0.8             &   1.19 -- 1.03      \\  
\em{J}                  &   2011-04-30        & 463             &   120              &   1.4             &   1.04 -- 1.12      \\  
\em{H}                  &   2011-01-24        & \ldots          &   180              &   1.6             &   1.01 -- 1.00      \\  
\em{K$_{\rm{s}}$}         &   2011-01-22        & \ldots          &    90              &   0.9             &   1.01 -- 1.38      \\  
\hline
\end{tabular}
\end{table*}

\subsection{Data Reduction}
\label{sec:Reduction}

The data were reduced in IRAF following the methods detailed in \cite{Bruursema2014} and the steps outlined in the WHIRC Reduction Manual (Joyce 2009)\footnote{http://www.noao.edu/kpno/manuals/whirc/}.  Briefly, an array linearity correction was performed using the WHIRC task \texttt{wprep.cl}, and the \texttt{mscred} package was used to remove the pupil ghost from the bias-subtracted flat-field. The pupil ghost template and pupil removal mask were downloaded from the  WIYN-WHIRC NOAO website\footnote{https://www.noao.edu/kpno/manuals/whirc/WHIRC.html}. Each individual science image was then sky-subtracted and flat-fielded using a median-filtered sky frame. 
Finally, bad pixels were removed using standard bad pixel masks and the IRAF \texttt{fixpix} routine.

\begin{figure*} 
\centering
\includegraphics[width=6.7in]{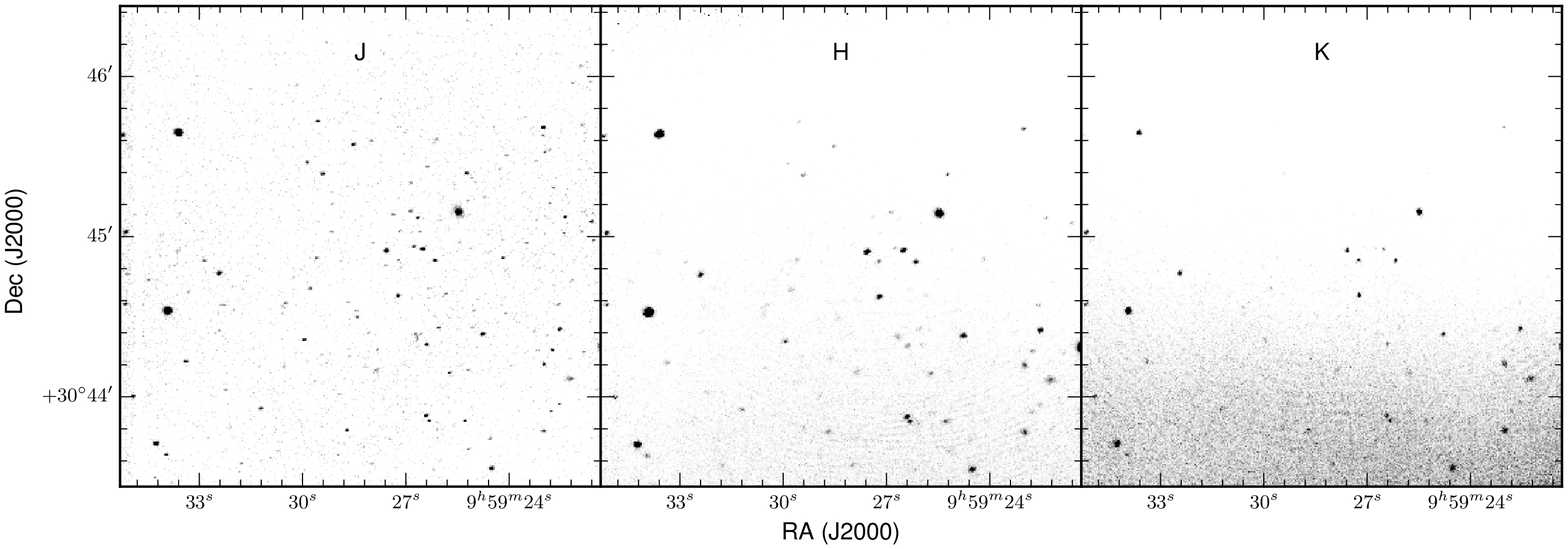}
\includegraphics[width=6.7in]{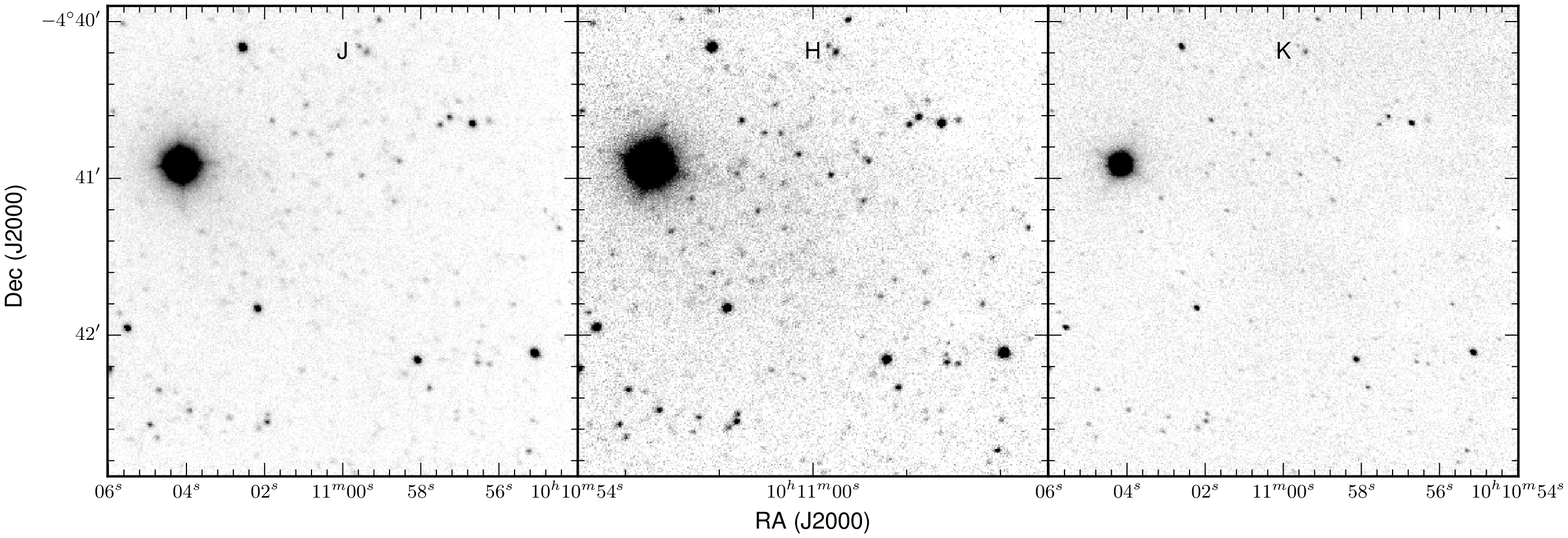}
\caption{WHIRC \emph{JHK}$_{s}$ maps of Leo A (top) and Sextans A (bottom). North is at the top, and east is to the left.}
\label{fig:galImages}
\end{figure*}

To determine the astrometry for Sextans A and Leo A, the PyRAF script \texttt{starfind} was used to find the positions of the brightest stars in each field. These stars were then compared to the positions of stars detected in the Two Micron All-Sky Survey (2MASS; \citealt{Skrutskie2006, Cutri2004}) to find the common stars in the image. Using these 2MASS stars as reference stars, the PyRAF packages \texttt{xyxymatch} and \texttt{geomap} were used to find and correct for position offsets in the frames so that the astrometry of the reduced images to match that of the 2MASS images. Field distortions in the images were corrected using the \texttt{geomap} and \texttt{geotrans} tasks in IRAF. Final mosaicked images in each filter for the galaxies were produced using {\sc Montage}\footnote{http://montage.ipac.caltech.edu/index.html}. The WHIRC maps of Sextans A and Leo A are shown in Fig.~\ref{fig:galImages}.

\subsection{Photometry}
\label{sec:Phot}

Source extraction, aperture photometry, and related functions were performed with {\sc astropy} and its recently developed subsidiary package {\sc photutils}\footnote{http://photutils.readthedocs.io/en/stable/} \citep{Astropy2013}. 

The {\sc DAOstarfinder} function was used to identify point sources in the images. Sharp and round limits were implemented in order to obtain a reliable list of point sources; this cut effectively removes cosmic ray hits and image artifacts, and minimizes the number of extended sources and resolved background galaxies from the source catalog.
Aperture photometry was performed on the resulting source list for each image. The radius of apertures was determined using the predicted full width half maximum of the image resolution, and then adjusted, based on the photometric quality. 
An outer annulus was used to determine the frame background level, and varied with each image proportionally to the area of the inner aperture.

Preliminary catalog cuts were implemented before flux calibration. Sources with error $\geq$65 percent or with flux $<$0 were eliminated, as they: (a) have a poor signal-to-noise ratio, (b) are background subtraction artifacts, or (c) a non-detection. The remaining sources are considered to be high-confidence point sources with reliable flux measurements.

Detector counts (ADU) of our good sources were calibrated into Janskys (Jy) via known 2MASS sources \citep{Skrutskie2006, Cutri2004} in the images.
With many of our images relying on 4--10 calibration-stars, a systematic conversion error of 2.5\% was added to the calibrated flux to account for scatter in the resulting corrections. However, the relative photometry accuracy within the images is better, as it only depends on the S/N of the source and the Poisson noise.

\subsection{Description of the Catalogue}

\subsubsection{The WHIRC Point-Source Catalog}
\label{sec:InterMatch}

The point source catalogue detection statistics are summarised in Table~\ref{table:catalogue_stats}. Table~\ref{tab:catalogue} lists the 779 sources included in the final point-source catalogue for both of the galaxies in our sample. For a source to be included in the full catalogue we require it to be detected in at least two bands or at two epochs with high-confidence. We consider a source to be a match if their centroids are within a 1'' radius. If a source was detected over multiple epochs we present the mean flux and uncertainty for the combined epochs. The  WHIRC high-reliability catalogue only includes sources detected in two or more WHIRC bands. This is less complete than the full catalogue but has a more stringent acceptance criteria.

\begin{table}
\begin{center}
\caption{Detection statistics}
\label{table:catalogue_stats}
\begin{tabular}{lcc}
   & & \\  
\hline
\hline
Wavebands                      &  Leo A   &  Sextans A  \\
\hline
Total Detected                 &  287     &  492        \\  
\emph{J, H {\rm\&} K}$_{s}$    &   38     &  176        \\  
 $J$ \& $H$                    &   98     &  88        \\  
 $J$ \& $K_{\rm{s}}$            &    1     &  24        \\  
 $H$ \& $K_{\rm{s}}$            &    3     &  11        \\
High-reliability catalogue   &   140    &  299        \\ 
WHIRC Variables              &   50     &  30         \\  
\hline
\end{tabular}
\end{center}
\end{table}

\begin{table*} 
  \centering
  \small
\caption{Catalog of Point Sources for Leo A and Sextans A}
\label{tab:catalogue}
 \resizebox{\linewidth} {\height}{    \setlength\tabcolsep{3pt}\begin{tabular}{lcccccccccccccccccc}
 & & & &   & & & &     & & & & & &     & & & &  \\
  \hline
  \hline
Galaxy & ID & $\mathrm{RA}$ &$\mathrm{Dec}$ & $U$ & $B$ & $V$& $I$  &$J$ & $\sigma J$ &$H$ &$\sigma H$  &$K$ &$\sigma K$ &  [3.6] & [4.5] & [5.8] & [8.0] & Variable \\ 
\hline
LeoA & 0 & 149.8928 & 30.7250 & -99.99 & 23.72 & 22.48 & 21.19 & 19.19 & 0.04 & 19.10 & 0.07 & -99.99 & -99.99 & 19.73 & 19.64 & -99.99 & -99.99 & 0 \\
LeoA & 1 & 149.8593 & 30.7257 & -99.99 & 23.44 & 22.34 & 21.06 & 19.31 & 0.04 & -99.99 & -99.99 & -99.99 & -99.99 & 20.12 & 19.04 & -99.99 & -99.99 & 0 \\
LeoA & 2 & 149.8522 & 30.7259 & -99.99 & 24.57 & 23.98 & -99.99 & 17.07 & 0.08 & 16.87 & 0.09 & 15.89 & 0.10 & 18.06 & 17.76 & 16.30 & -99.99 & 0 \\
\hline\\
\end{tabular}}
\end{table*}

\begin{figure} 
\centering    
\includegraphics[clip=true,width=3.2in,  trim=1cm 0cm 0cm 0cm]{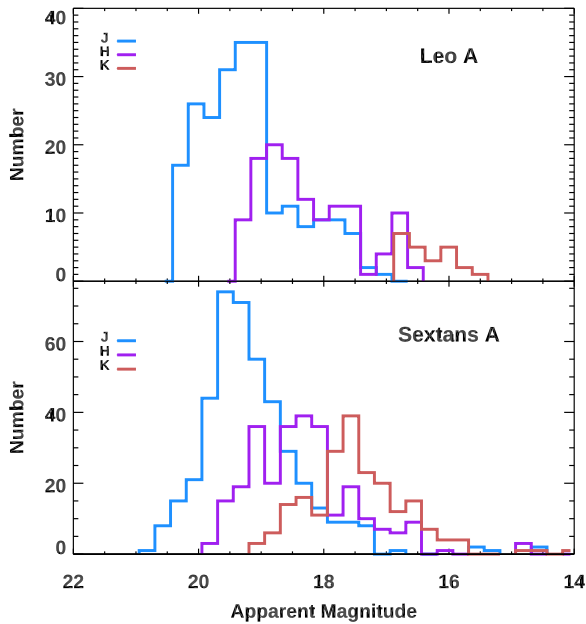} 
\caption[]{Histogram of the number of sources detected in the three wave bands as a function of magnitude.}
  \label{fig:lumFn}
\end{figure}

Figure~\ref{fig:lumFn} shows a histogram of the number of sources detected in the \emph{JHK}$_{s}$  photometric bands as a function of magnitude.
The faint source detection limit for the K$_{s}$ data of Leo A is almost two magnitudes brighter than for Sextans A due to the high sky background. 
Although less deep this data is useful in constraining the SEDs of the brightest sources in the galaxy.

\subsubsection{Ancillary Photometric Data}
\label{sec:Match}

In order to fully investigate the nature of the sources in our near-IR catalogue; we match our source catalogue to all available photometric catalogs for Leo A and Sextans A on a purely positional basis. A matching distance of 3'' was adopted, and if multiple matches occured within a single catalogue, only the closest match was kept.
The ancillary data was obtained from the Hubble Space Telescope ({\em HST}) \citep{Bianchi2012}, the Large Binocular Telescope (LBT) \citep{Bellazzini2014}, the Subaru Prime Focus Camera (Suprime) \citep{Stankute2014}, and {\it Spitzer Space Telescope} \citep{Boyer2009, Boyer2015a, Boyer2015b}. 
These catalogues are not contemporary with the WHIRC data. However, they can aid in source classification and place constraints on the spectral energy distribution (SED) from the optical to the mid-IR. Table~\ref{tab:otherPhot} lists each ancillary catalogue, the photometric bands and the number of sources in common with our master catalogue. 

 Most sources ($>$90\%) are matched to at least one of these ancillary catalogs. At longer wavelengths the number of matches decreases as dust-enshrouded objects are rarer than stellar photospheres.
Of the matched sources, nine stars in Sextans A and two stars in Leo A were identified as AGB variable star candidates by \cite{Boyer2015a, Boyer2015b}. 
Our catalogue also includes the two known planetary nebulae \citep{Magrini2003} in these galaxies.
 WHIRC sources without optical counterparts are rare, however, the large differences in angular resolution and survey depth increases the likelihood for mismatches compared to the WHIRC-IRAC matches. We discuss these objects further in Section~\ref{sec:Discussion}.

\begin{table*} 
\begin{center}
\caption{Number of matches between the WHIRC Catalogs and other Photometric Catalogs}
\label{tab:otherPhot}
\begin{tabular}{lcccc}
 &&    &&\\ 
\hline
\hline
Survey                           &  Filters       &  Telescope       &   Leo A         &  Sextans A  \\   \hline                                                                                                                                        
 WHIRC - High reliability        &   $JHK_{\rm{s}}$ &  WIYN           &     140         &  299        \\    DUSTiNGS; \cite{Boyer2015a}     &  [3.6], [4.5]  &  {\em Spitzer}  &     125         &  272        \\    DUSTiNGS Var; \cite{Boyer2015b} &  [3.6], [4.5]  &  {\em Spitzer}  &       2         &    9        \\    \cite{Boyer2009}                &  [5.8], [8.0]  &  {\em Spitzer}  &      58         &  120        \\    \cite{Bianchi2012}              &  $U,B,V,I$     &  {\em HST}      &  \ldots         &  177        \\    \cite{Bellazzini2014}           &  $g,r$         &  LBT            &  \ldots         &  292        \\    \cite{Stankute2014}             &  $B,V,I$       &  Subaru         &     138         & \ldots      \\    \hline
\end{tabular} 
\end{center}
\end{table*}

\section{Analysis and Results}
\label{sec:Results}

\subsection{Variable stars}
\label{sec:var}

\begin{figure} 
\centering
\includegraphics[clip=true,width=3.3in, trim=1cm 0cm 0cm 0cm]{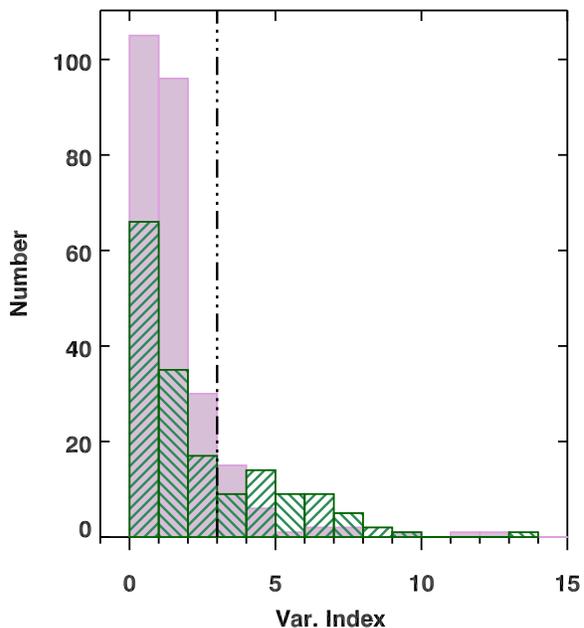}
\caption[]{The variability index distribution of all the sources detected in at least two epochs for Leo A (green lines) and Sextans A (pink). We consider a source to be variable if the $J$-band variability index $>$ 3 (dashed-dotted line).}
\label{fig:var}
\end{figure}

Evolved stars pulsate, and are variable on timescales of hundreds of days. These pulsations levitate atmospheric material, leading to the formation of dust grains. Thus stars with long period variations can show a significant IR excess. 
Our data examines source variability for three epochs, separated by 82, 267 and 363 days for Sextans A, and two epochs separated by 463 days for Leo A. 
We identify large-amplitude variables from our J-band data using the variability index defined by \cite{Vijh2009}.
This variability index takes into account the error-weighted flux difference between epochs. Figure \ref{fig:var} shows the variability index distribution for Sextans A and Leo A. Following \cite{Vijh2009}, sources with $|{\rm Var}| > 3 $  are classified as variable. This conservative threshold (adopted from the LMC's IR stellar populations) corresponds to a $J$-band amplitude of a least 0.25 mag for our sample. We find 30 variables in Sextans A and 50  variables in Leo A meet this criteria (Table~\ref{tab:catalogue}).

Our total sample has  11 sources in common with the DUSTiNGS variable catalogue. Of these, we identify three as variable using our J-band criteria.
The majority of these candidate variable stars are bright, with well-characterized errors. Variability indices for fainter stars are less constrained, as small changes in amplitude will be masked by the photometric errors,  this effects the identification of the lowest amplitude variable stars in our catalogue, and those stars significantly obscured by dust \citep{Whitelock2017}. This bias is more pronounced for Sextans A  due to its greater distance.

\begin{figure*} 
\centering
\includegraphics[clip=true,width=5.2in]{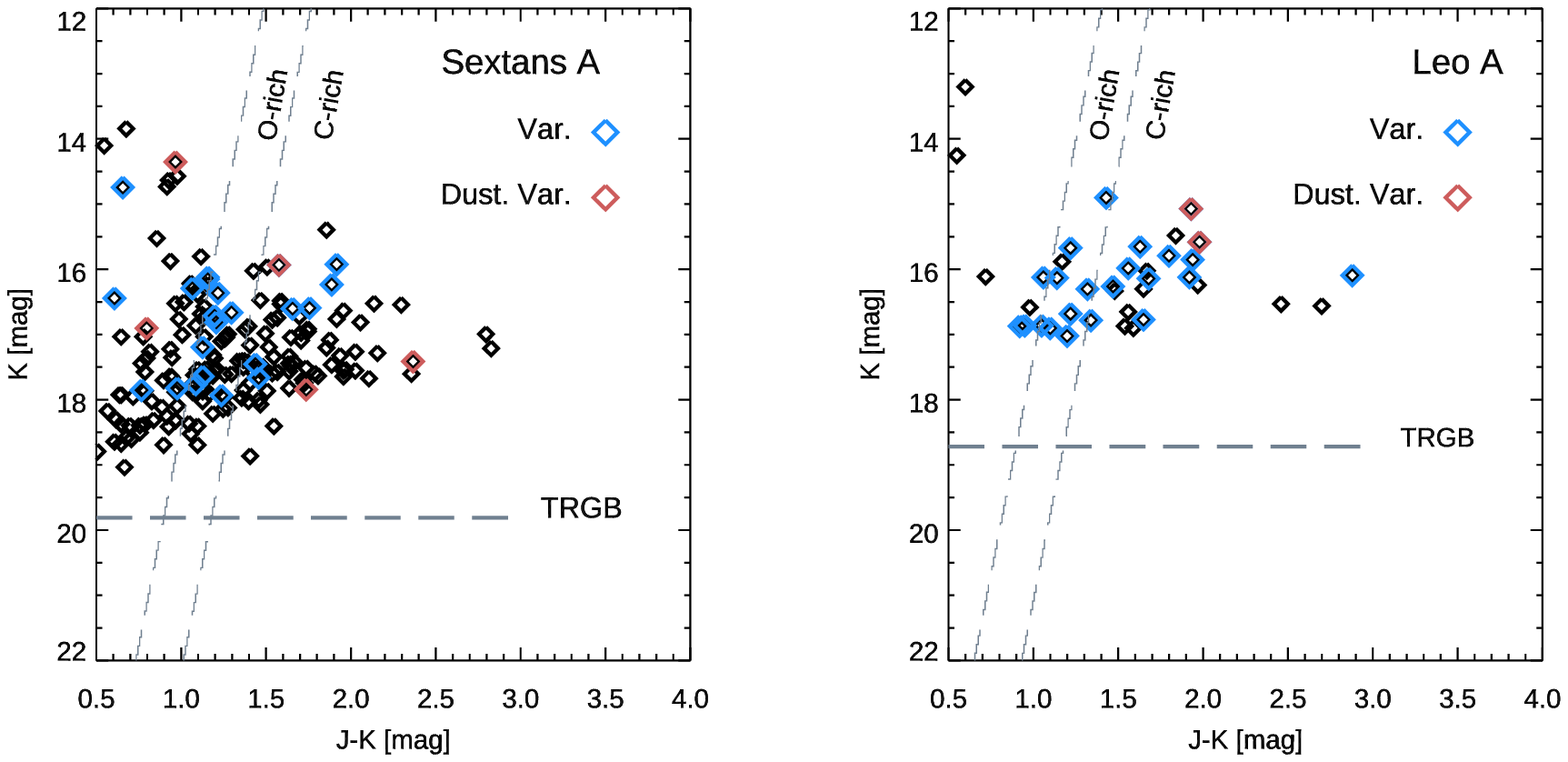}
\includegraphics[clip=true,width=5.2in]{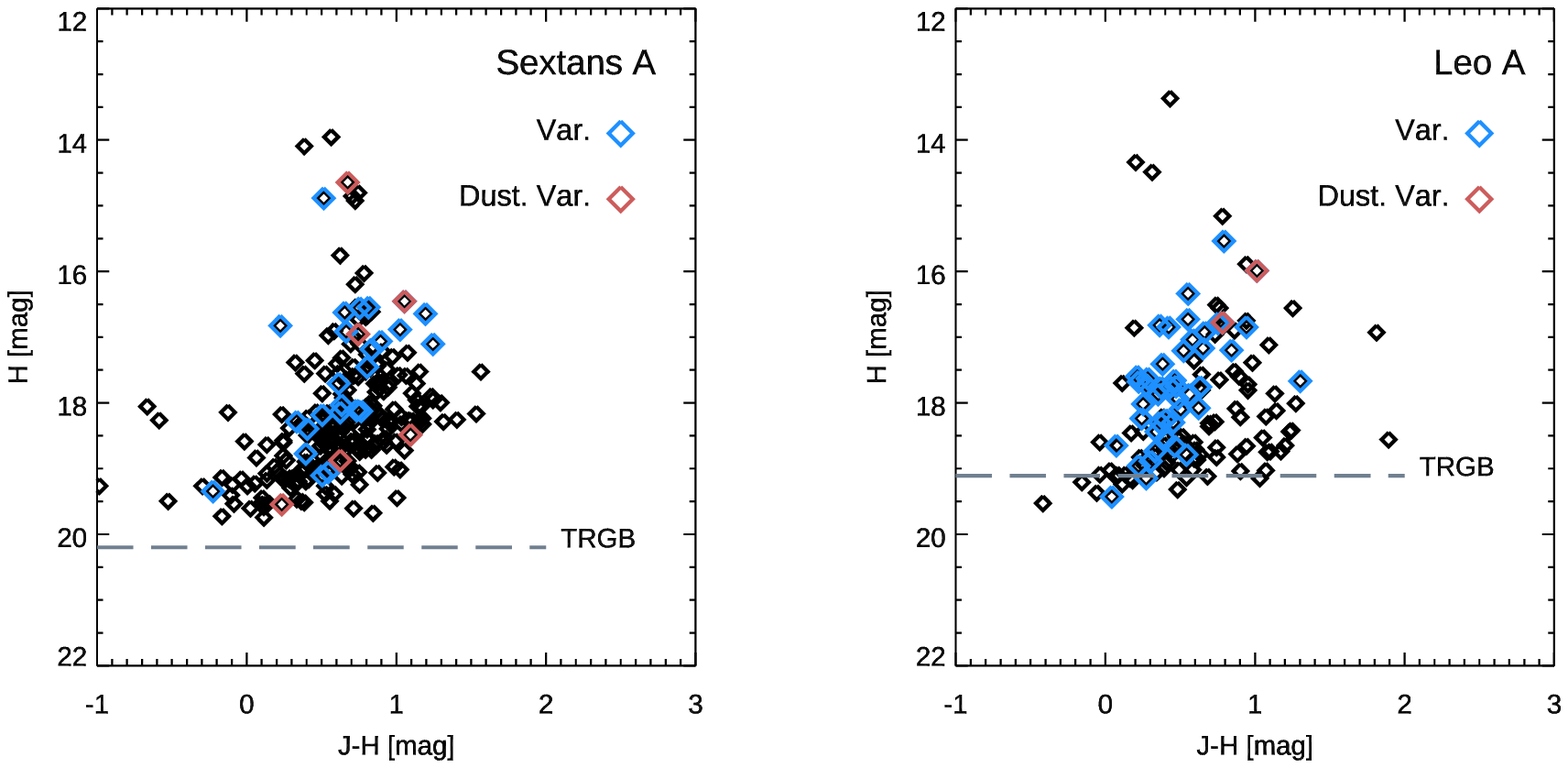}
\includegraphics[clip=true,width=5.2in]{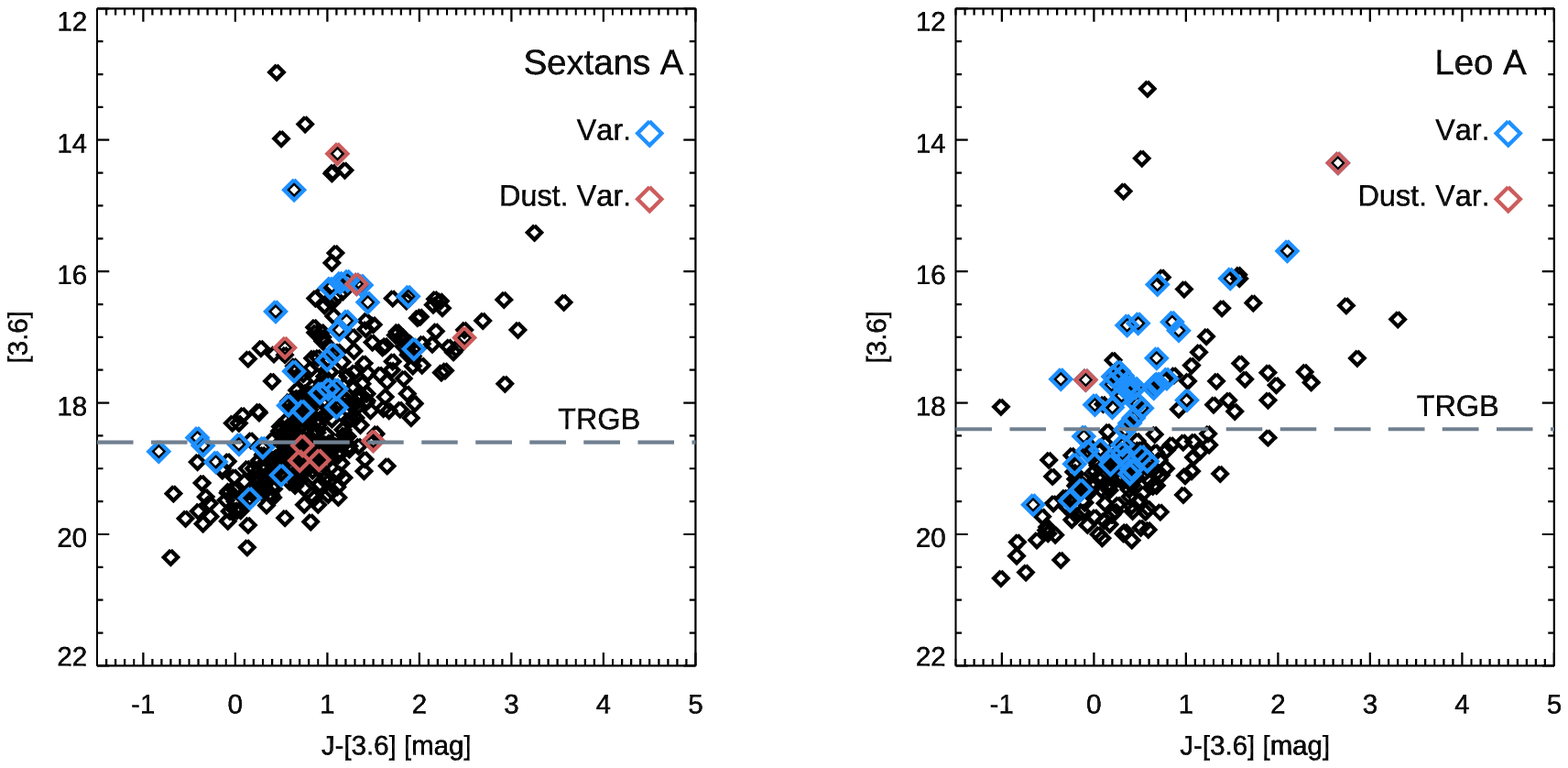}
\vspace{0cm} 
\caption[]{Near- and mid-IR colour-magnitude diagrams of stars in Sextans A (left) and Leo A (right), with the average magnitudes of variable star candidates over-plotted. Upper panel: $J-{K}_{{\rm{S}}}$ vs. ${K}_{{\rm{s}}}$ diagram with cuts adopted from \cite{Cioni2006} and \cite{Boyer2011} (diagonal dashed lines). Middle panel: $J-H$ vs. $H$ diagram. Lower Panel: $J-[3.6]$ vs. $[3.6]$ diagram. In each CMD, the estimated location of the TRGB is shown as horizontal line. Note that the number of sources in the $J-H$ and the $J-K_{s}$ CMDs are not the same for Leo A, due to the limited sensitivity of the K$_{s}$ data.}
  \label{fig:CMD1}
\end{figure*}

\subsection{Colour-Magnitude Diagrams}
\label{sec:CMDs}

In order to classify the variables we use colours and magnitudes to separate AGB stars from other sources. Near-IR colours are particularly useful for roughly discriminating between O- and C-rich AGB stars, and other stellar populations due to their distinctive spectral characteristics. O-rich stars have spectra dominated by VO, TiO and H$_2$O molecules, whilst stars with an excess of carbon in the atmosphere are characterized by CN and C$_2$. The CMDs constructed from the dereddened photometric measurements (see Section~\ref{sec:DPR}) are shown in Figure~\ref{fig:CMD1}. 

Sources are identified as TP-AGB stars or RSGs candidates if they are brighter than the RGB tip and redder than foreground objects.  The empirical relationship determined by \cite{Valenti2004}  was used to estimate the near-IR RGB tip in Leo A and Sextans A. Due to the low mass of these systems identifying other features in the CMDs is challenging. 
Instead, we adopt the cuts to the K$_{s}$ versus J$-$K$_{s}$ CMD used by \cite{Cioni2006} and \cite{Boyer2011} to separate the C-AGB and O-AGB stars, and identify dusty AGB stars (which experience molecular blanketing) as those with $J-[8.0]\gtrsim 3.4$ mag or $J-{K}_{{\rm{s}}}\gtrsim 2.2$ mag \citep{Blum2006, Boyer2011}. The location of these cuts, determined from the molecular content and opacity in the stellar photospheres, are shown in Figure~\ref{fig:CMD1}. In near-IR CMDs, O-AGB stars populate a narrow sequence extending to brighter magnitudes and redder colours, above the tip of the red giant branch (TRGB); carbon stars occupy a broad region to the right of the O-AGB sequence in the form of a red branch \citep{McDonald2012, Boyer2015c}.

The brightest sources in the CMDs are foreground stars and they have relatively blue colors. Using the {\sc trilegal} Galactic population synthesis code \citep{Girardi2005} we expect about 10 foreground stars brighter than $J=21$ mag in our Leo A field, and 30 foreground stars in Sextans A, all with $J-H < 0.6$ mag.

It should be noted that the variation in metallicity between the galaxies may result in changes to the IR colour.  The metallicity of the population affects the separation between the parallel lines shown in the top panel of Figure~\ref{fig:CMD1}. At higher-metallicity, O-rich AGB stars occupy a broader range of $J-K$ colours, at low-metallicity their colour distribution is narrower. Thus caution is necessary in the adopting these colour classifications as the location of the cuts are dependent on the metallicity of the galaxy and the extinction correction, limiting the classification accuracy.  As both Leo A and Sextans A have populations formed across a range in metalicities, this may affect the photometric chemical classification of approximately 20\% of the AGB stars in our sample.

For sources without $K_{s}$ band detections we use the $H$-band data to determine an initial class following the cuts proposed by \cite{Sibbons2012}. 
Stars with  $J-H < 0.6$ mag are most likely to be main-sequence stars or foreground dwarfs, and stars with  $J-H > $ 1.0 mag are classified as carbon stars. 
For completeness, we also use the $J-[3.6]$ versus [3.6] CMD (i.e. $J-[3.6] \gtrsim 3.1$ mag) suggested by \citet{Blum2006} to separate the O-rich, C-rich, and dusty AGB populations. This necessary as molecular blanketing due to dust strongly affects the near-IR spectral region.

For red sources, contamination from unresolved background galaxies may be significant. Focusing on the variable star population eliminates the majority of this contamination. Conversely given the SFHs of Sextans A and Leo A, contamination from massive young stellar objects ($M > 5\,M_{\rm \odot}$) is likely very small.
The vertical sequence seen in the $J-[3.6]$ CMD is populated by objects with a range in stellar effective temperatures (see Section~\ref{sec:HRdiagram}), and little IR-excess. 
In both galaxies, the number of very red objects is modest.
The results of the photometric classifications of evolved star candidates are listed in Table~\ref{table:photclass}.

\begin{table}
\small
\caption{Photometric Classification:  evolved star candidates}
\label{table:photclass}
\begin{tabular}{lcc}
\hline
\hline
Population                              &	Leo A     &	Sextans A      \\ 
\hline                                                
$J < TRGB_{[J]}$                         &      57         &    295             \\
O-rich AGB                              &      10         &    57              \\
C-rich AGB                              &      18         &    62              \\
Dusty AGB stars	                        &       3         &    18              \\
RSGs                                    &       1         &    8               \\
Foreground                              &       2         &    3               \\
\hline
\end{tabular}
\end{table}

\subsection{Hertzsprung-Russell diagrams}
\label{sec:HRdiagram}

In order to calculate the effective temperature  ($T_{\rm eff}$)  and bolometric luminosities ($M_{\rm bol}$) of the sources in Sextans A and Leo A, we used the SED-fitting code of \cite{McDonald2012, McDonald2017}. This code performs a $\chi^2$-minimization between the observed SED (corrected for interstellar reddening) and a grid of {\sc bt-settl} stellar atmosphere models \citep{Allard2011} which are scaled in flux to derive a bolometric luminosity. 

The {\sc bt-settl} spectral models are based on atomic and molecular data; for each galaxy we used literature values (given in Table~\ref{tab:galProp}) for the metallicity, distance, and interstellar reddening. 
A $\chi^2$-minimisation is performed, comparing the observed SED to a blackbody function. The derived $T_{\rm eff}$ and $M_{\rm bol}$ is then used as a starting point for an iterative $\chi^2$-minimisation against the interpolated stellar model atmosphere grid. In this way, a precise ($\pm$200 K) fit to the data is achieved. 
For cases where the SED is not well-sampled, uncertainties are higher. Furthermore in crowded fields, or areas with background nebulosity, luminosities may be more uncertain.

\begin{figure} 
\centering
\includegraphics[clip=true,width=3.1in, trim=1cm 0cm 0cm 0cm]{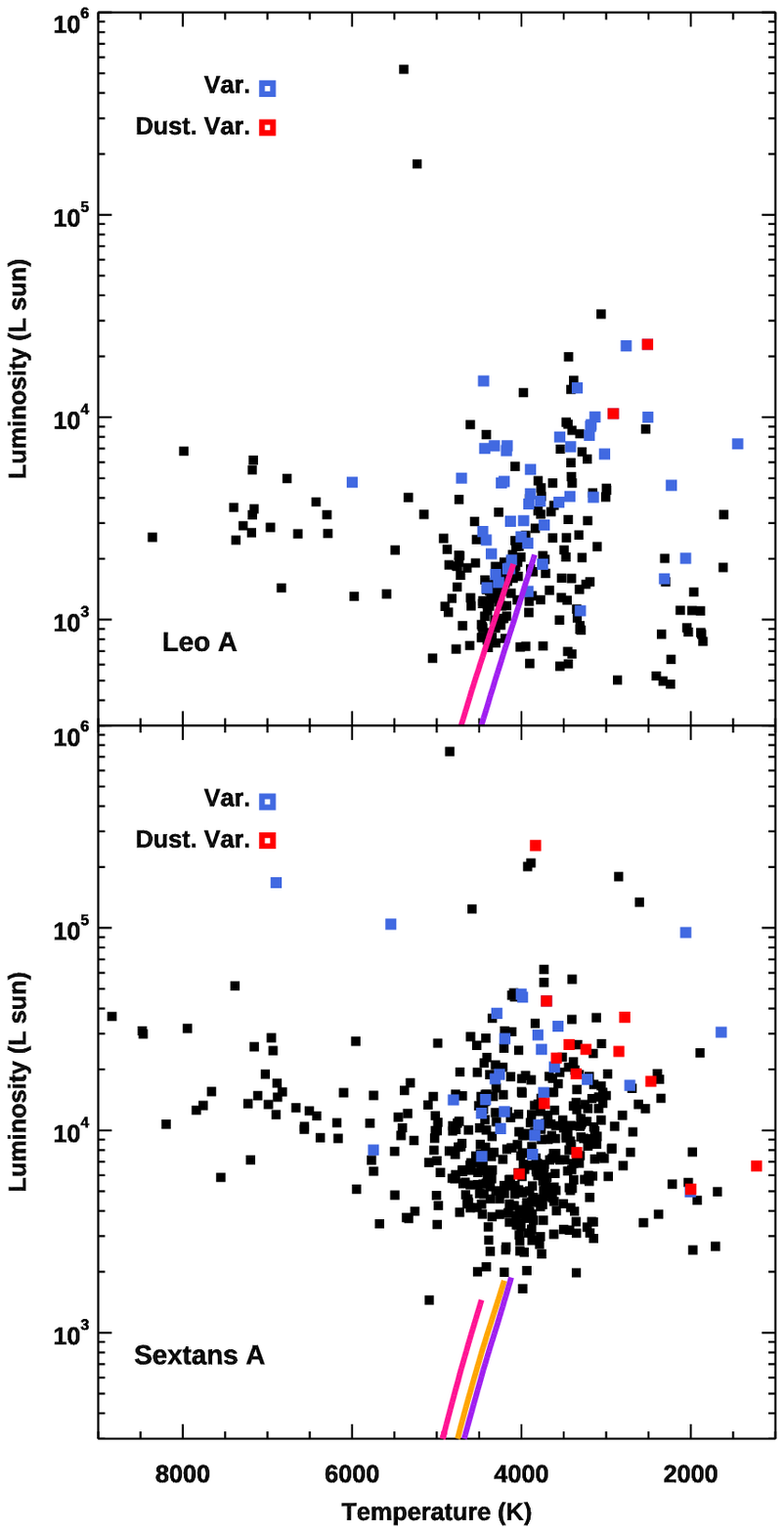} 
\vspace{0cm} 
\caption[HR diagram]{The Hertzsprung--Russell diagrams of Leo A (top) and Sextans A (bottom). WHIRC variable stars are shown in blue and DUSTiNGS variables in red.  Isochrones for [Fe/H] $= -1.4$ and $[\alpha/H]  = 0.4$, for a 3 Gy (pink) and 14 Gy (purple) population, calculated form the Dartmouth models \citep{Dotter2008} are overlaid on the Leo A data.  Isochrones of [Fe/H] $= -1.85$ and $[\alpha/H]  = 0.4$ for 3 Gy (pink), 8 Gy (orange)  and 14 Gy (purple) populations are shown for Sextans A. The top of the isochrones correspond to the RGB tip.
}
\label{fig:HR_diagram}
\end{figure}

Figure~\ref{fig:HR_diagram} shows the Hertzsprung--Russell diagram of Leo A and Sextans A we have produced using the derived effective temperatures and bolometric luminosities for all the stars in our field of view.
In Leo A, a concentration of stars on a well defined RGB is clearly visible, with the RGB tip at approximately 3800 K, 2500 ${\rm L}_{\odot}$. A small AGB population appears to continue onward from this, ascending to the top right of the diagram. To the left of the RGB is a population of hotter stars, these are likely to be foreground stars, main sequence stars or point sources with insufficient IR data constraining the SED. However, some post-AGB stars, OB stars, Herbig Ae/Be stars, and certain kinds of binary stars, could be present at in this region at higher luminosities. In Leo A, sources cooler than 3500--3000 K, with luminosities less than 3000 ${\rm L}_{\odot}$ are likely to be background galaxies, although some galaxies may also reside amongst the RGB and AGB stars. 

In Sextans A there is a large amount of scatter in the H--R diagram, and the precise location of the RGB is difficult to define. The scatter is due to a more populated main-sequence and giant-branches, combined with worse photometric accuracy. 
Foreground stars are visible in towards the top left of the H--R diagram, whilst background galaxies lie at cooler temperatures and low luminosities.

In an attempt to separate the RGB and AGB stars from the rest of the population in Sextans A we over-plot our variable star candidates and compare their location to that of the RGB in Leo A.
When comparing the two galaxies, it is clear that in Sextans A we are sensitive to only the brightest AGB stars and our H--R diagram is therefore incomplete, even at the tip of the RGB, whilst in Leo A we reach a fainter giant branch population. There may also be a difference between the bright AGB populations.
In Sextans A, the scatter in the AGB population may be a result of a higher concentration of carbon stars due to the lower bulk metallicity of the galaxy.

\subsection{Dust-production rates}
\label{sec:DPR}

Mass-loss rates for every evolved star in Leo A and Sextans A can be computed by fitting their broadband spectral energy distributions (SEDs) with radiative transfer models. This allows us to characterize the cumulative dust production in these galaxies, and constrain the evolutionary status and chemical type of the individual stars.

To fit the SEDs of all the stars in our catalogue, we use the grid of dusty RSG and AGB models \citep[{\sc grams};][]{Sargent2011, Srinivasan2011}. These models were computed using the {\sc 2Dust} radiative transfer code \citep{Ueta2003}, and stellar photosphere models for O-rich \citep{Kucinskas2005, Kucinskas2006} and C-rich \citep{Aringer2009} AGB stars.
The O-rich dust is modeled using astronomical silicates \citep{Ossenkopf1992}, and for carbon stars, a mixture of amorphous carbon (amC) and 10\% silicon carbide (SiC) was used.

To model the stars, we use SEDs constructed from the mean flux at each wavelength and we correct for interstellar reddening using the \cite{Cardelli1989} extinction law with $R_{\rm v} =  3.1$ mag.
Uncertainties in the optical and near-IR photometric data were increased in quadrature using representative LPV amplitudes, to account for stellar pulsations \citep[see][]{Srinivasan2016}. This allows us to fit SEDs for which a precise phase correction could not be determined. 

As the {\sc grams} models are computed for stars in the LMC, the model fluxes must be scaled before performing the fitting. After correcting for distance and extinction, an optimal scaling factor ($\eta$) was determined by minimising the difference between the observed and model fluxes. 
We then perform a ${\chi }^{2}$ minimisation between the models and the data to obtain the best-fit parameters, which include the dust-production rate (DPR), and chemical type.
 The ${\chi }^{2}$ value is defined by:
\begin{equation}
\chi ^2_\nu = \frac{1}{N}\sum _i \frac{(f_{{\rm obs}_i} - f_{m_i})^2}{\sigma _i^2},
\end{equation}
where $f_{{\rm obs}_i}$ and $f_{m_i}$ are the observed and model flux in the ith band with error bar $\sigma _i$, and $N$ is the total number of data points.
As a scaling factor ($\eta$) has been applied to the models, the model luminosity and wind parameters must also be adjusted. The luminosity scales by $\eta$  and the mass-loss rate by the factor $\sqrt{\eta}$, following equation (2) of \cite{Groenewegen2006}.

To avoid degeneracies in the model fits, we compare the best fit parameters from the carbon- and oxygen-rich models separately, adopting an approach similar to that of \cite{Srinivasan2016}. Here we consider all models that satisfy $\chi^2 -\chi_{\mathrm{best}}^{2} < 2N_{\mathrm{data}}$, to be acceptable, where $ \chi_{\mathrm{best}}^{2}$ is the reduced ${\chi }^{2}$~value of the best-fitting model and $N_{\mathrm{data}}$ is the number of photometric data points. The uncertainty in each best-fitting parameter is determined from the median absolute deviation of these models.
The chemical classification (carbon or oxygen rich) is established from the model with the smallest overall ${\chi }^{2}$, with a chemical uncertainty determined from the ${\chi }^{2}$ ratio of the best-fit carbon- and oxygen-rich models.

\begin{figure*} 
\centering
  \includegraphics[clip=true,width=2.4in, trim=0.3cm 1cm 0cm 0cm]{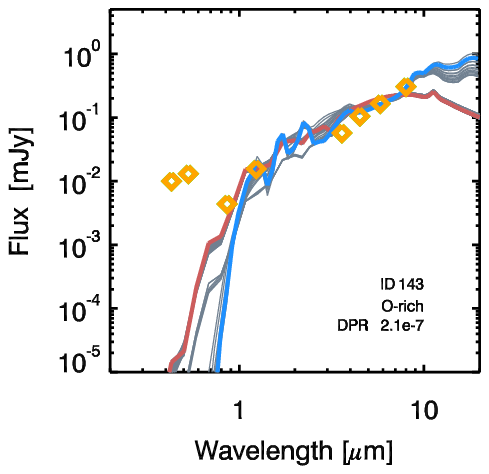}
  \includegraphics[clip=true,width=2.0in, trim=1.3cm 1cm 0cm 0cm]{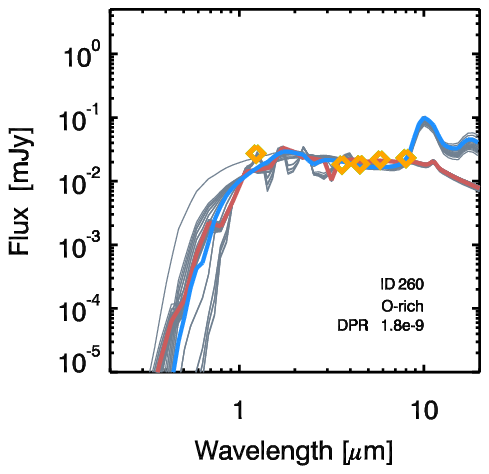}
  \includegraphics[clip=true,width=2.0in, trim=1.3cm 1cm 0cm 0cm]{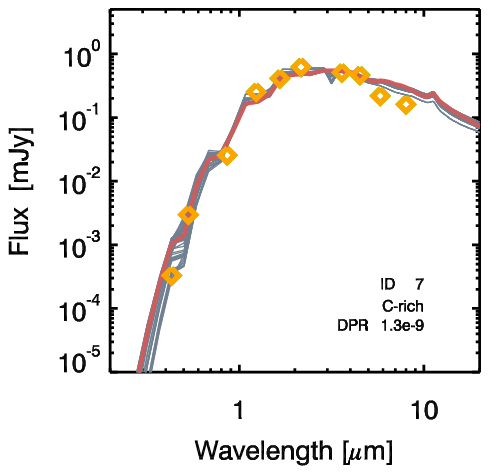}
  \includegraphics[clip=true,width=2.4in, trim=0.3cm 0cm 0cm 0.5cm]{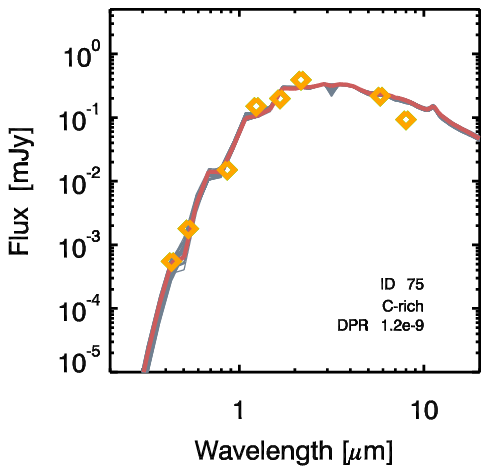}
  \includegraphics[clip=true,width=2.0in, trim=1.3cm 0cm 0cm 0.5cm]{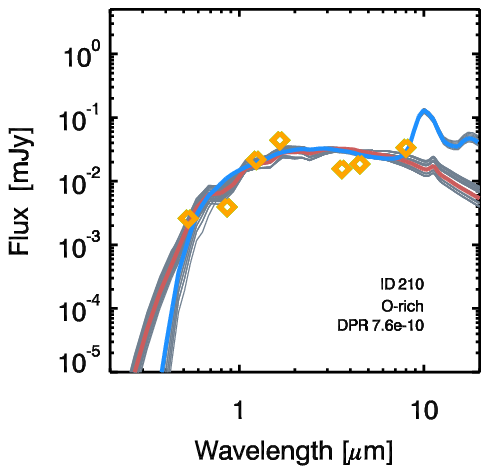}
  \includegraphics[clip=true,width=2.0in, trim=1.3cm 0cm 0cm 0.5cm]{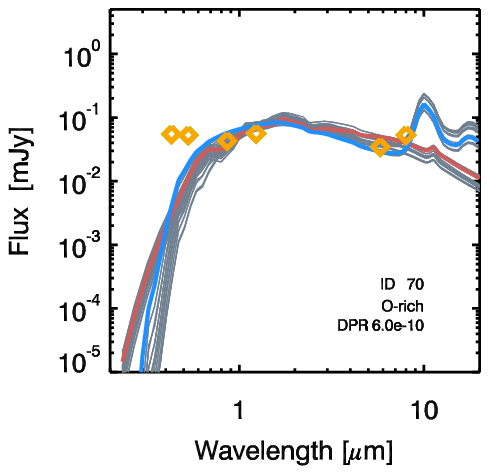}
  \caption[]{The SED and best-fit {\sc grams} models for the six sources with the highest dust-production rates in Leo A. The orange diamonds are the observed photometry corrected for interstellar reddening, the best fit oxygen-rich model is shown in blue, and the best fit carbon-rich model in red. The grey lines show the range in the acceptable model fits for each source. For the reddest sources, the optical flux can be affected by deep molecular absorption bands.     The anomalous optical photometry, for some sources, is most likely due to a mismatch between catalogues or due to variability. LeoA-143 was identified as a candidate PNe from H$\alpha$ observations by \cite{Magrini2003}.
  }
\label{fig:leoA_fits}
\end{figure*}

\begin{figure*} 
  \centering
  \includegraphics[clip=true,width=2.4in, trim=0.3cm 1cm 0cm 0cm]{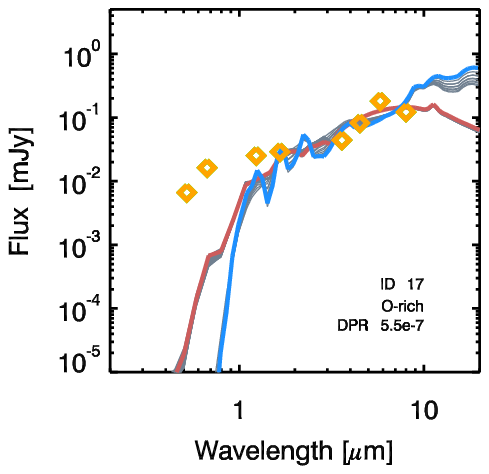}
  \includegraphics[clip=true,width=2.0in, trim=1.3cm 1cm 0cm 0cm]{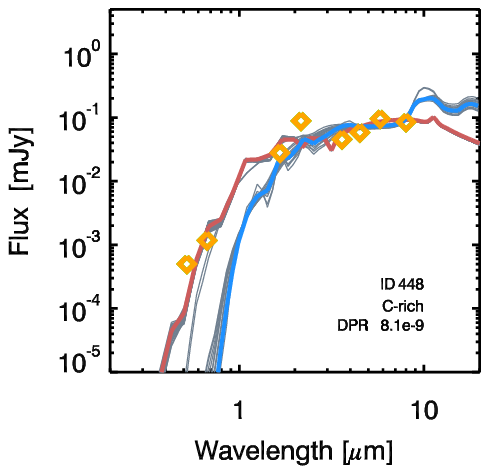}
  \includegraphics[clip=true,width=2.0in, trim=1.3cm 1cm 0cm 0cm]{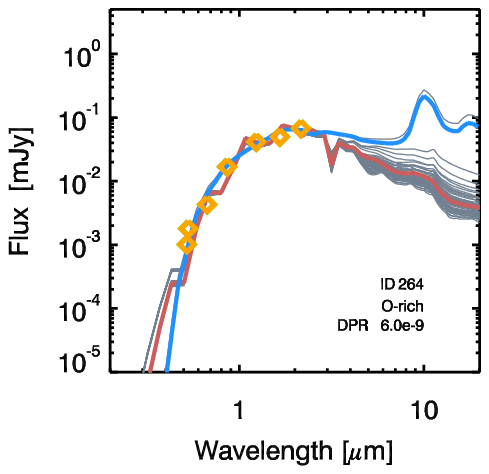}
  \includegraphics[clip=true,width=2.4in, trim=0.3cm 0cm 0cm 0.5cm]{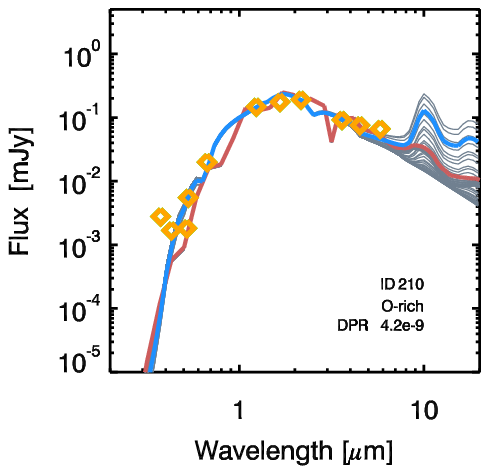}
  \includegraphics[clip=true,width=2.0in, trim=1.3cm 0cm 0cm 0.5cm]{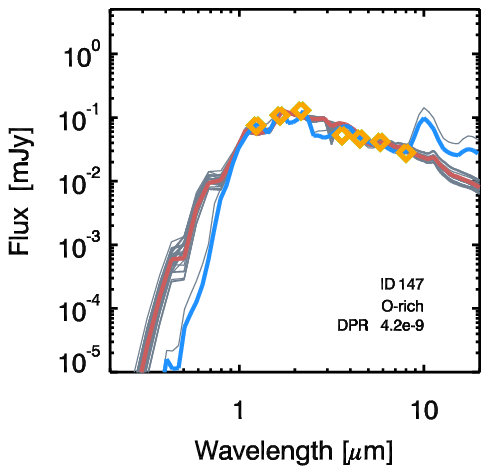}
  \includegraphics[clip=true,width=2.0in, trim=1.3cm 0cm 0cm 0.5cm]{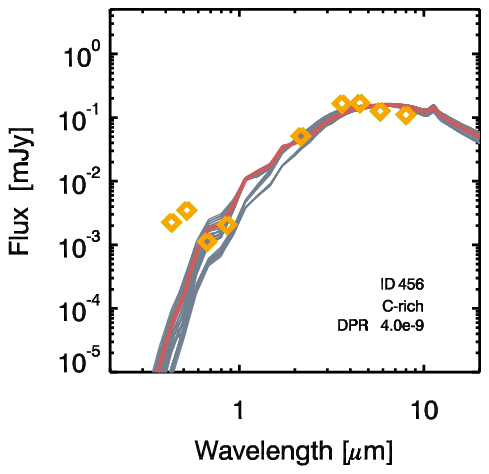}
\caption[]{The SED and best-fit {\sc grams} models for the six sources with the highest dust-production rates in Sextans A. Colours are the same as Figure~\ref{fig:leoA_fits}.}
\label{fig:sextansA_fits}
\end{figure*}

Figures~\ref{fig:leoA_fits} and \ref{fig:sextansA_fits} show the SEDs and best-fitting models for the  six dustiest sources in each galaxy. In both figures, the range of acceptable fits for each source is shown in grey, and where appropriate, both the best fit carbon and oxygen rich model. In instances where only one chemistry provides an acceptable fit to the data, only that chemical type is shown. For the reddest sources, and SEDs with high ${\chi }^{2}$ values, we carefully inspect the SED and model output to verify the quality of the fitting procedure. In most instances, objects with a high ${\chi }^{2}$ have SEDs which are inconsistent with a stellar object; they are disjointed, with a poorly defined shape (produced from a mismatch between catalogs) or have too few valid data points. We consider these SEDs to be invalid and discarded their fits from the dust-budget estimates. The results of the model-fits for 162 stars in Leo A and 281 stars in Sextans A are listed in Table~\ref{table:DPR}.

\begin{table*}
  \small
  \centering
\caption{Luminosities and dust-production rates of the 443 evolved stars with valid fits in Leo A and Sextans A.}
\label{table:DPR}
\begin{tabular}{lcccccc}
&&&&&&      \\     
\hline
\hline
Galaxy & WHIRC\_ID  & $T_{\rm eff}$  & $M_{\rm bol}$ & DPR & $\sigma$DPR & GRAMS Chem. \\
       &            &   (k)        &             & (${\rm M}_{\odot} \, {\rm yr}^{-1}$)  & (${\rm M}_{\odot} \, {\rm yr}^{-1}$) &  \\ 
\hline
LeoA & 	3 & 	3800 & 	$-$4.1 & 	5.0E$-$11 & 	1.1E$-$11 & C  \\
LeoA & 	4 & 	4100 & 	$-$3.7 & 	2.3E$-$13 & 	4.8E$-$12 & O  \\
LeoA & 	7 & 	2500 & 	$-$6.1 & 	1.3E$-$09 & 	4.3E$-$10 & C  \\
LeoA & 	9 & 	4200 & 	$-$3.2 & 	1.4E$-$13 & 	2.9E$-$12 & O  \\
\hline
\end{tabular}
\end{table*}

\subsubsection{Uncertainties in dust-production rates }

\begin{figure} 
\centering
\includegraphics[clip=true,width=3.2in, trim=1.3cm 1.5cm 0cm 0cm]{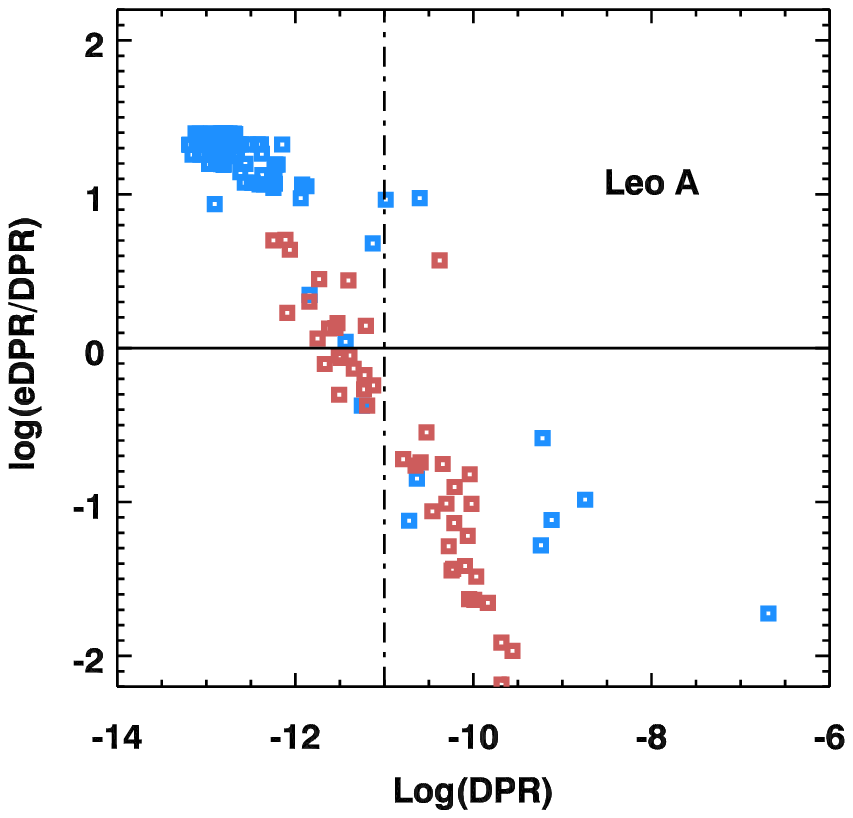}
\includegraphics[clip=true,width=3.2in, trim=1.3cm 0cm 0cm 0.7cm]{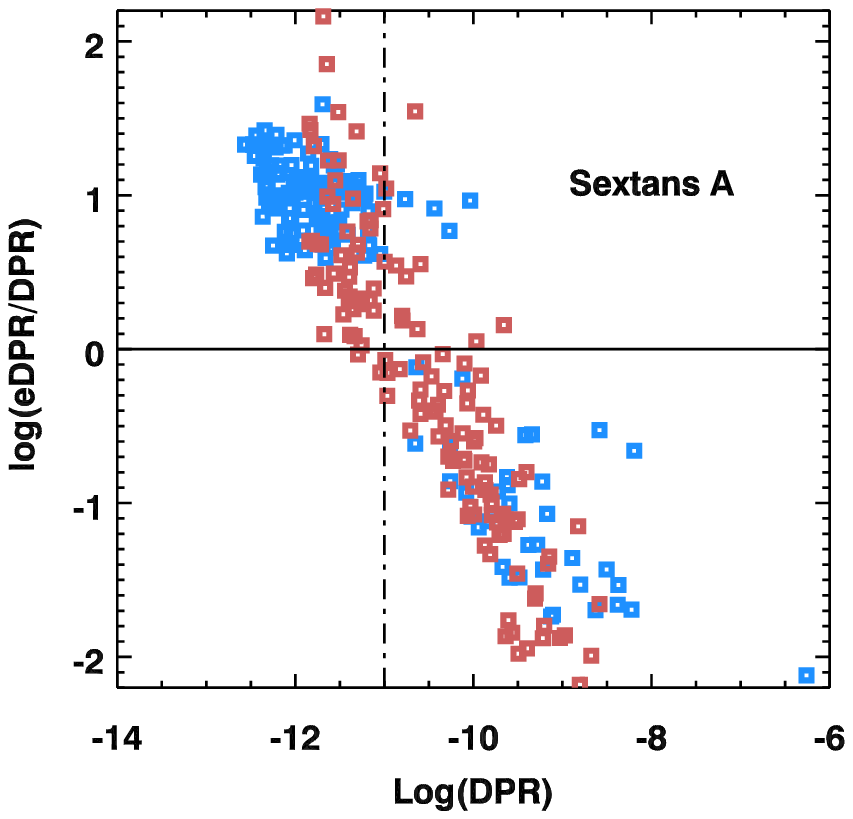}
\caption[]{The relative uncertainty in the dust-production rate as a function of the dust-production rate for Leo A (top) and Sextans A (bottom). Oxygen-rich stars are blue and carbon-rich stars are red. Sources to the left of the dashed line are consistent with a mass-loss rate of zero. Stars in the lower-right quadrant are producing dust. Only these sources were used to determine the global dust-mass injection into the ISM of each galaxy. }
\label{fig:errorDPR_fit}
\end{figure}

In Figure~\ref{fig:errorDPR_fit} we compare the relative uncertainty in the dust-production rate as a function of the dust-production rate for our sample. Sources with dust-production rates of $\log({\rm DPR})<-11$ have large uncertainties and are consistent with stellar photospheres with no apparent circumstellar excess due to dust. In both Sextans A and Leo A, the majority of the stars detected  ($\sim$70\%) are low-DPR sources hence do not contribute to the dust budget of the galaxy.

When comparing the dust-production rate between galaxies, we have assumed that there is no change in the dust properties with metallicity, and that the wind expansion velocity is constant. A fixed expansion velocity is unlikely if radiation pressure on dust is the primary accelerant of the wind. This has been observed by \cite{Marshall2004}, \cite{Groenewegen2016} and \cite{Goldman2017} who found a decrease in expansion velocity for stars with comparable properties at lower metallicity. Furthermore, \cite{Jones2012, Jones2014, McDonald2010} have shown that the composition of oxygen-rich dust species has a dependence on metallicity. These systematically increase the absolute uncertainty in both the individual dust-production rates and the global dust injection rates (see Table 2 of \citealt{McDonald2011} for a detailed analysis of the inherent uncertainties in dust-production rates). However, the star-to-star variations are much smaller.

\begin{figure} 
\centering
\includegraphics[clip=true,width=3.2in, trim=1.3cm 1.5cm 0cm 0cm]{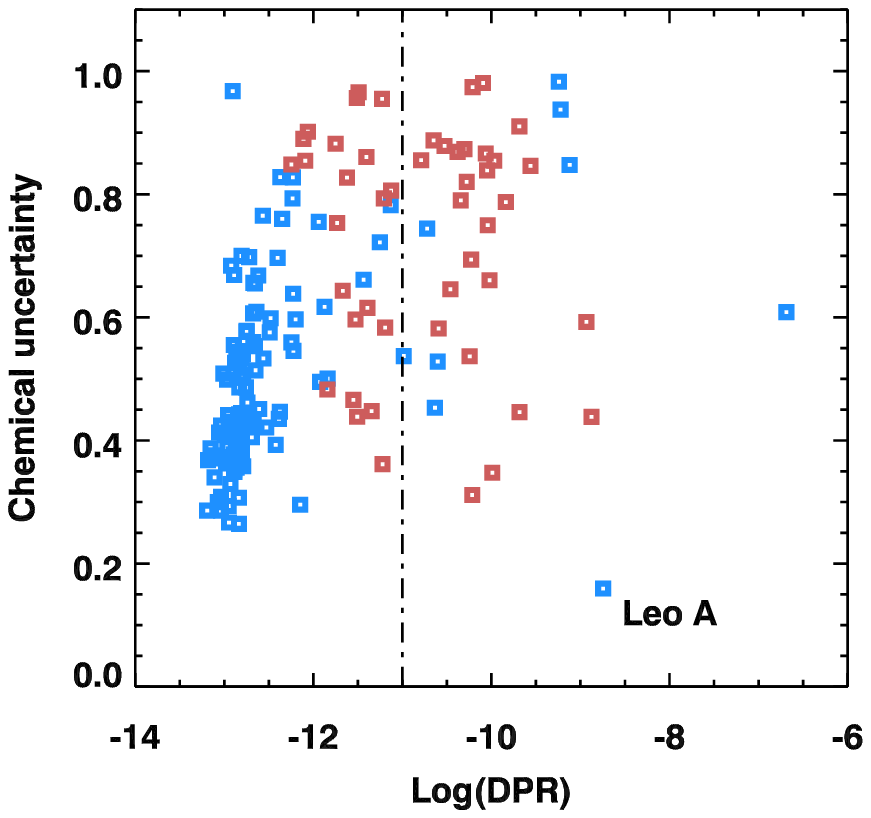}
\includegraphics[clip=true,width=3.2in, trim=1.3cm 0cm 0cm 0.7cm]{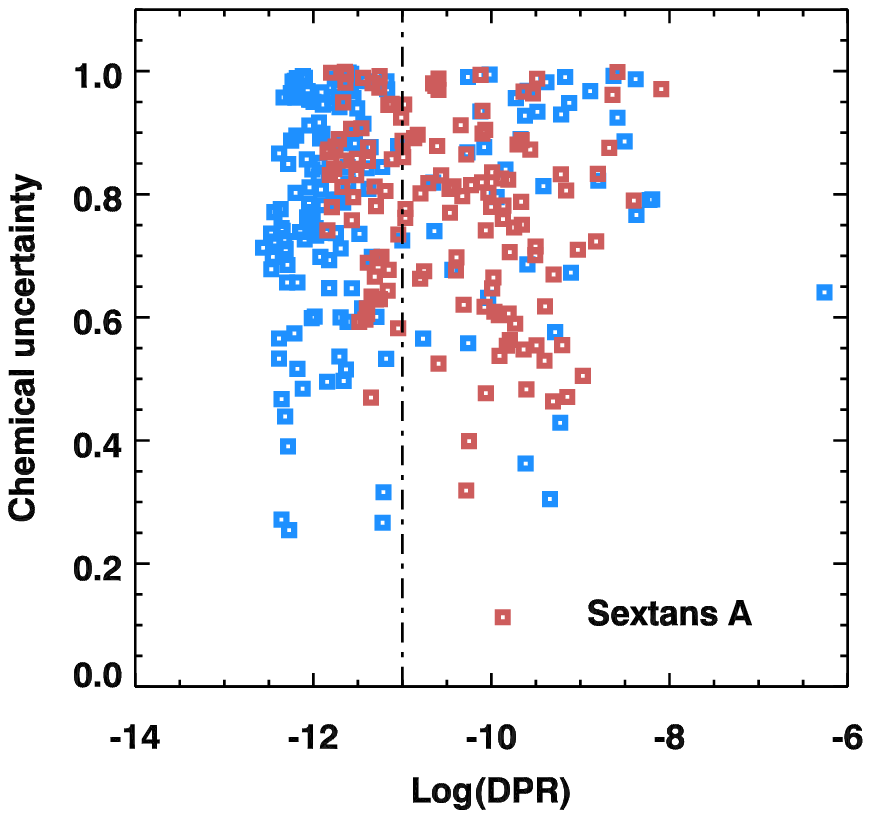}
\caption[]{The ratio of the ${\chi }^{2}$~values for the best-fitting models from the carbon- and oxygen-rich model grids for each source. A low value ($<0.7$) corresponds to a high-confidence chemical classification.}
\label{fig:chemConfidence}
\end{figure}

The chemical classification determined from the best-fit {\sc grams} models also has a large influence on the uncertainty in the dust-production rates. Low-confidence chemical classifications, where both the O- and C-rich models produce almost an equally good fit to the data, can have dust-production rate estimates that differ by over an order of magnitude. 
Figure~\ref{fig:chemConfidence} shows the uncertainty in the chemical classification. 
The majority of the sources in our sample with $\log ({\rm DPR}) > -11$ were deemed to have a large chemical uncertainty, with a value close to unity. This is a consequence of a lack of data at mid-IR wavelengths, necessary for breaking the degeneracy in dust chemistry.

\section{Discussion}
\label{sec:Discussion}

\subsection{Late-type stellar populations of Leo A and Sextans A}

\begin{figure} 
\centering  
\includegraphics[clip=true,width=3.2in, trim=0.5cm 0cm 0cm 0.0cm]{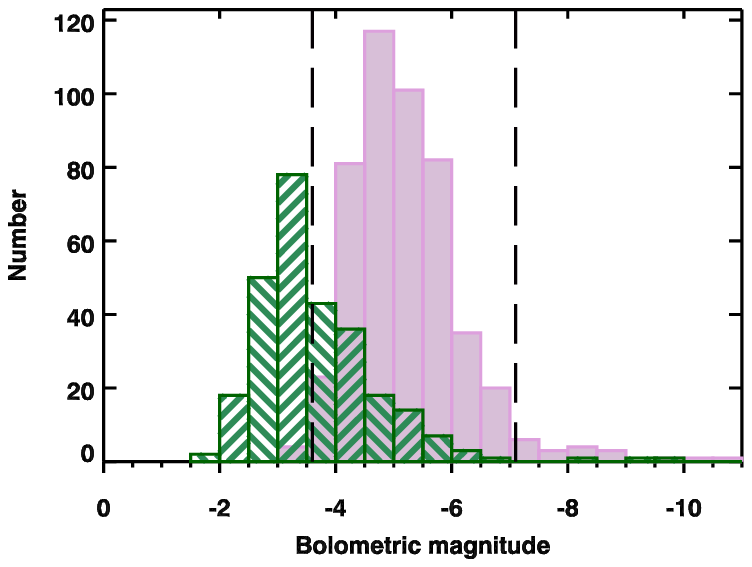} 
\includegraphics[clip=true,width=3.2in, trim=0.5cm 0cm 0cm 0.0cm]{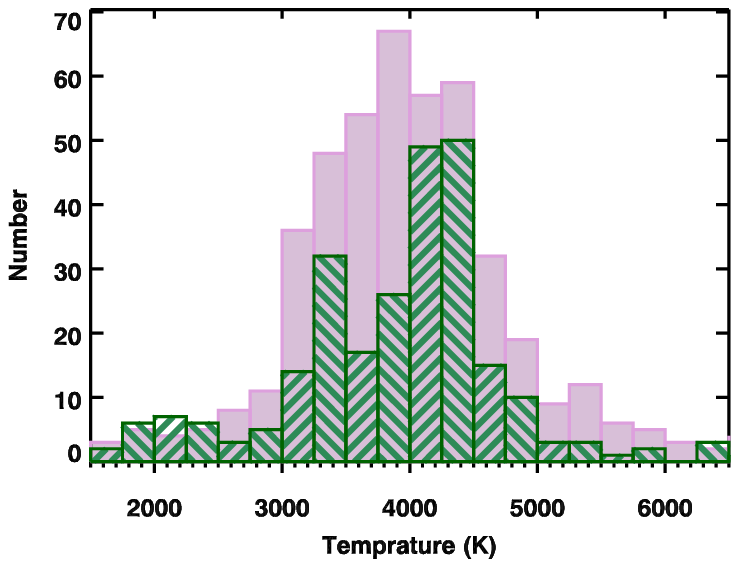} 
\caption[]{Temperature and luminosity distributions of Leo A (green lines) and Sextans A (pink). The dashed lines in the top panel mark the TRGB at $M_{\rm bol} =-3.6$  \citep{Ferraro2000} and the classical AGB limit at $M_{\rm bol} =-7.1$.}
\label{fig:temDist}
\end{figure}

The differences in the temperature and luminosity distributions between the stellar populations of Leo A and Sextans A can be clearly seen in Figure~\ref{fig:temDist}. In Leo A there is a significant decline in source density towards brighter magnitudes at $M_{\rm bol} =-3.6$, which corresponds to the TRGB \citep{Ferraro2000}. Stars below the TRGB are of low mass ($M \sim0.8-2\,M_{\rm \odot}$), likely from a population-aged between 1 and 13 Gyr. The depth of the photometry for Sextans A is not sufficient to reach the TRGB; here the luminosity function has a rapid decline towards fainter magnitudes at $M_{\rm bol} = -4.5$.  Our reported AGB population should therefore be considered a lower limit.
The discontinuity near $M_{\rm bol} = -4.5$, may also occur if there was a significant increase in the star-formation rate (SFR) a few hundred Myr ago.

Leo A has a blue population of giants that have low circumstellar extinction and populate fainter magnitudes, which is consistent with RGB or early-AGB stars. Leo A has only a few red sources that appear to be evolving on the TP-AGB, these stars are bluer than those in Sextans A.
The presence of carbon stars in the CMDs provides confirmation of an intermediate-age population in both Leo A and Sextans A. Only intermediate-mass AGB stars ($\sim$2--4 ${\rm M}_{\odot}$; \citealt{Marigo2007, Ventura2010}) can become carbon-rich, although the transition luminosity is lower in metal-poor populations.
Sextans A appears to have a higher relative fraction of carbon stars compared to Leo A (Table~\ref{table:DPR}), and its evolved stellar population (above the TRGB) is more luminous.

In both galaxies there are a few stars with bolometric magnitudes brighter than the traditional AGB luminosity limit of $M_\mathrm{bol}=-7.1$ \citep{Wood1983}.  These bright sources represent a small population of RSGs or hot-bottom-burning AGB stars (with $M > 8\,M_{\rm \odot}$), whose ages are consistent with a recent increase in the star-formation rate $\sim$1 Gyr ago \citep{Dolphin2005, Cole2007}.
The discrepancy in the size of the AGB populations unlikely due to the difference in stellar mass of the host galaxy. More likely it is due to differences in both the mean metallicity and star-formation history.

\subsection{Dust Production}

Stars with $\log ({\rm DPR})>-11$ have a measurable IR excess due to dust. 
There are 32 stars in Leo A and 101 stars in Sextans A which meet this criteria, with a low-moderate relative uncertainty in the DPR.
As the uncertainty in the chemical classification is high for most sources (Figure~\ref{fig:chemConfidence}), we compute four values for the total dust production rate: 1) using the best-fit models irrespective of dust chemistry; 2) assuming all stars are O-rich; 3) assuming all stars are C-rich; and 4) only including stars identified as variable. Table~\ref{tab:DPRvalues} lists the total dust-production rates for each galaxy. These estimates exclude any source with a relative uncertainty in the DPR $>$0 (see Figure~\ref{fig:errorDPR_fit}).

\begin{table*}
  \centering
\caption{The global dust-production rates for Leo A and Sextans A}
\label{tab:DPRvalues}
\begin{tabular}{lcc|cc}
\hline
\hline
Population           & \multicolumn{2}{c}{Leo A}  & \multicolumn{2}{c}{Sextans A}      \\
                     & Total DPR & Median DPR      &   Total DPR  & Median DPR            \\
                     & (${\rm M}_{\odot} \, {\rm yr}^{-1}$)  & (${\rm M}_{\odot} \, {\rm yr}^{-1}$)  & (${\rm M}_{\odot} \, {\rm yr}^{-1}$)  & (${\rm M}_{\odot} \, {\rm yr}^{-1}$)  \\
\hline                                                
Best-fitting chemical type     &  (8.2 $\pm$ 1.8)   $\times 10^{-9}$    &      8.1 $\times 10^{-11}$  &   (6.2 $\pm$ 0.2)  $\times 10^{-7}$  &    1.8 $\times 10^{-10}$  \\
If all O-rich        &  (6.5 $\pm$ 2.8)   $\times 10^{-9}$    &      6.2 $\times 10^{-10}$  &   (6.3 $\pm$ 1.5)  $\times 10^{-7}$  &    4.5 $\times 10^{-10}$  \\ 
If all C-rich        &  (5.0 $\pm$ 1.5)   $\times 10^{-9}$    &      8.1 $\times 10^{-11}$  &   (4.4 $\pm$ 1.3)  $\times 10^{-8}$  &    1.1 $\times 10^{-10}$  \\
 Variables only      &  (2.9 $\pm$ 0.8)   $\times 10^{-9}$    &      2.1 $\times 10^{-10}$  &   (5.7 $\pm$ 0.1)  $\times 10^{-7}$  &    1.6 $\times 10^{-9}$  \\
\hline
\tablenotetext{}{Best-fitting chemical type: 25 C-rich and 6 O-rich in Leo A, and 68 C-rich and 33 O-rich in Sextans A.}
\tablenotetext{}{ The total DPR in Sextans A should be considered a lower limit, as we do not detetect AGB stars to the tip of the RGB.} \end{tabular}
\end{table*}

In both galaxies the total dust input into the ISM is dominated by one or two stars. In Leo A the star with the highest dust-production rate is LeoA-143, with a DPR of (2.1 $\pm$ 0.1) $\times 10^{-7}$ ${\rm M}_{\odot} \, {\rm yr}^{-1}$. This is two orders of magnitude greater than the second-highest dust-production rate in Leo A. This object has a rising SED and was identified as a candidate PNe from H$\alpha$ observations by \cite{Magrini2003}, hence we exclude this source from the global dust-budget estimate.

In Sextans A, the two sources with the highest dust-production rates are IDs 17 and 448 (SSTDUSTG J101059.43-044249.3 and J101058.05-044304.1, with (5.5 $\pm$ 0.1) $\times 10^{-7}$ ${\rm M}_{\odot} \, {\rm yr}^{-1}$ and  (8.1 $\pm$ 3.9) $\times 10^{-9}$ ${\rm M}_{\odot} \, {\rm yr}^{-1}$, respectively). These also have rising SEDs, however both these sources have been identified as variable x-AGB candidate by DUSTiNGS \citep{Boyer2015b}. 
Our fitting routine classified  SextansA-17 as an oxygen-rich AGB-star; it is responsible for 89 per cent of the global dust production of Sextans A. If this source is instead carbon rich, then its dust-production rate is significantly lower: (1.2 $\pm$ 0.3) $\times 10^{-8}$ ${\rm M}_{\odot} \, {\rm yr}^{-1}$. If SextansA-17, is confirmed to be an oxygen-rich AGB star, then it conclusively shows that individual metal-poor ($[{\rm Fe/H}] \sim -1.85$ to $-1.40$) evolved stars can produce significant amounts of dust, and contribute substantially to the enrichment of the ISM in their host system.

\begin{figure} 
\centering
\includegraphics[clip=true,width=3.2in, trim=1.3cm 1.5cm 0cm 0cm]{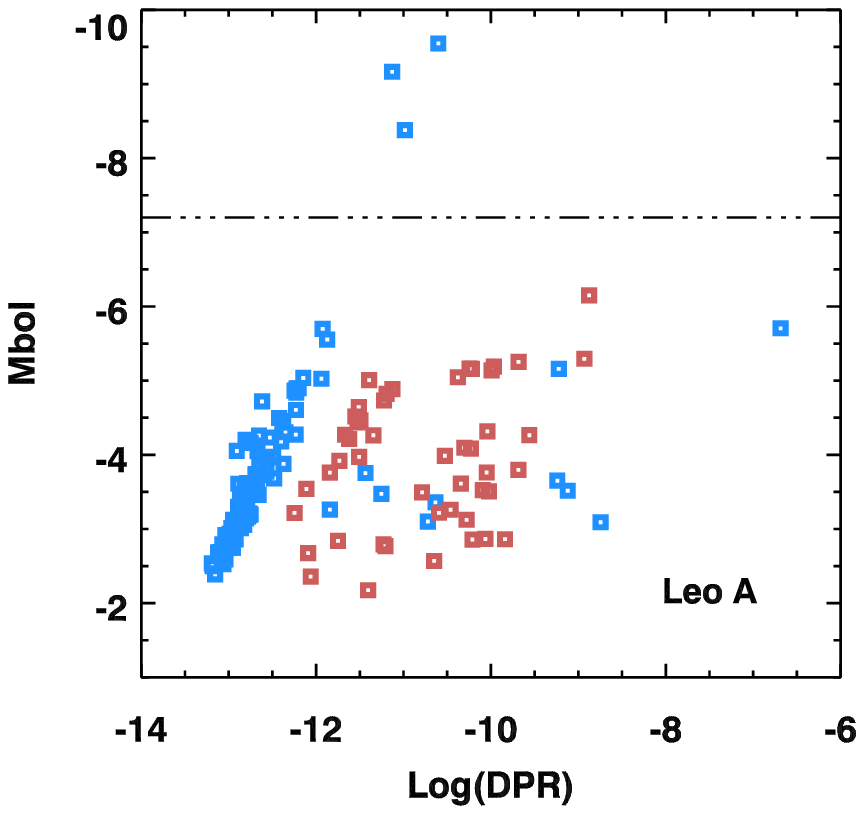}
\includegraphics[clip=true,width=3.2in, trim=1.3cm 0cm 0cm 0.7cm]{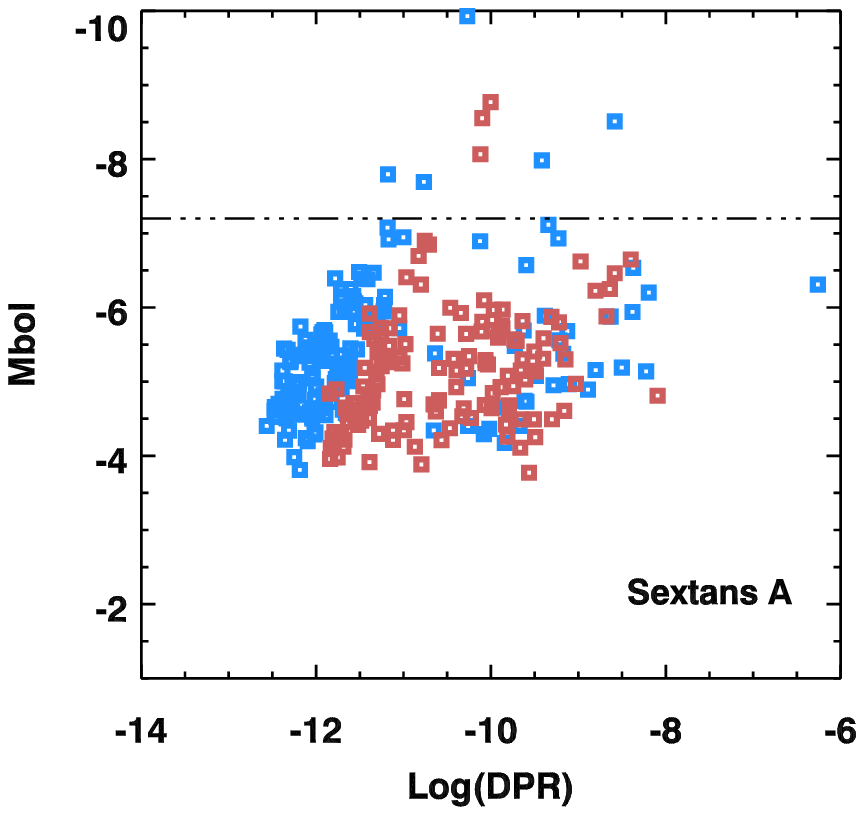}
\caption[]{The bolometric magnitude as a function of the dust-production rate for Leo A (top) and Sextans A (bottom). The horizontal line marks the classical AGB luminosity limit, all other symbols are as in Figure~\ref{fig:errorDPR_fit}.}
\label{fig:MbolvsDPR}
\end{figure}

Figure~\ref{fig:MbolvsDPR} compares the dust-production rate to a source's bolometric magnitude. For the stars with a circumstellar excess, $\log ({\rm DPR})>-11$, no trend is seen between mass-loss rate and luminosity. This is surprising, as both luminosity and dust-production rate should increase during the stars' evolution on the AGB, as expected from most mass-loss prescriptions \citep[e.g.][]{Bloecker1995, Vassiliadis1993}.
A similar effect was seen in the LMC by \cite{Riebel2012}.
Three luminous carbon stars are observed in Sextans A which are in excess of the maximum AGB luminosity limit, however all three have low-confidence chemical classifications. If, they are indeed carbon-rich they are unlikely to be AGB stars associated with the galaxy and are presumably foreground stars.
All stars brighter than $M_{\rm bol}= -9.0$ are considered foreground sources, and are excluded from global DPR estimates.

The most extreme AGB stars, producing the largest amount of dust are near to the end of their AGB evolution. The RGB-tip is at approximately $M_{\rm bol} \sim -3.6$ mag, TP-AGB stars should all be brighter than this limit \citep[e.g.,][]{Rosenfield2014}.  Consequently, our data represents a near-census of the TP-AGB stars in Leo A and we can separate the dust-producing AGB stars from the larger RGB population. In Sextans A, we do not reach the RGB tip as we are limited by sensitivity, so our dust production value for this galaxy should be considered a lower limit.

\subsubsection{Comparison of global dust-production}

Sextans A is at a lower metallicity than Leo A yet the cumulative dust-production rate of Sextans A is significantly higher than for Leo A, even when accounting for the difference in stellar mass between the galaxies. 
This result is consistent with \cite{Boyer2015b}, who estimated the rate of dust production by AGB stars in Leo A to be 5.2 $\times 10^{-9}$ $M_\odot \,{\rm yr}^{-1}$, and Sextans A to be 9.3 $\times 10^{-8}$ $M_\odot \,{\rm yr}^{-1}$.
However, our results disagree with the trends observed from the evolved-star dust budgets of the Magellanic Clouds \citep{Srinivasan2016}. Here the SMC is observed to have consistently lower DPRs than the LMC; this is thought to be a due to the lower mean metallicity of the galaxy.

As the LMC and SMC are closer,  have well characterized evolved star populations
and have SEDs that are better sampled at IR wavelengths than Leo A and Sextans A, the integrated mass-loss of the Magellanic Clouds is less likely to be affected by stochastics from a few rare objects that are experiencing strong mass loss, contamination from other source types, and by systematic errors in the SED fitting.

The main source of uncertainty in our dust-budget estimates are due to the ambiguous evolutionary status of a small number of sources, rather than the chemical uncertainty of the dust.
This is because the dust-production rates determined with carbon-star models are, on average, higher than with oxygen-rich models, for individual stars with a small IR-excess. Conversely, the oxygen-rich models return higher dust-production rates than the carbon-rich models for redder stars.

For both galaxies the median dust-production rates calculated by \cite{Boyer2015b} is about an order of magnitude higher than our estimates.
This is expected as the \cite{Boyer2015b} values only includes variable stars redder than $[3.6]-[4.5] > 0.1$. Whereas our calculations include all stars with a circumstellar excess, the majority of which are producing dust at low rates and are not in the super-wind phase of evolution. However the same general trend is seen with the median dust output showing no effects due to metallicity.

Approximately a quarter of all the stars that are producing dust at rates of  $\log ({\rm DPR})>-11$ are best fit with oxygen-rich models. This is unexpected since the abundance of silicon atoms at these metallicities is thought to be too low to form silicate dust grains  \citep{DiCriscienzo2013, Zhukovska2013, Schneider2014}, whereas carbon stars can still produce dust, due to the progressive enrichment of carbon atoms in the surface regions via dredged up from the nuclosynthesis in the core \citep{Karakas2009}. Thus we would expect to see more carbon-rich dust at lower metallicities, and almost no silicate dust.

As noted earlier, the constraints on the dust chemistry are poor, however, if these AGB stars are forming silicate dust then this suggests that these stars may be more metal-rich, and hence younger and of higher mass, than the rest of the AGB population in Leo A and Sextans A. Alternatively, this dust might be nucleating in a circumbinary disc, or that silicate dust may form via an alternate reaction mechanism to metal-rich stars, from silicon atoms synthesized at the end of the RSG evolutionary phase \citep{Weaver1980}. The mass-loss process, in these metal-poor stars, may also be fundamentally different to the pulsation-enhanced radiation-driven wind \citep[e.g.][]{Hoefner2008} we might naively expect.
To verify these results, observations with {\em JWST} in the mid-IR would place significant constraints on both the dust chemistry and the stellar type.

\section{Summary}  
\label{sec:conclusion}

\emph{JHK}$_{s}$ images taken with the  WIYN telescope at Kitt Peak are used to investigate the evolved star content of the Local Group galaxies Leo A and Sextans A. These observations have allowed us to characterize the near-IR stellar populations using variability, and colour-magnitude diagrams to identify evolved stars. In Sextans A only stars above the TRGB are detected. Matching our data to broadband optical and mid-IR photometry we determine luminosities, temperatures and dust-production rates for each star via SED fitting and radiative-transfer modeling. 
We identify 32 stars in Leo A and 101 stars in Sextans A with $\log ({\rm DPR})>-11$, confirming that metal-poor stars can form substantial amounts of dust. Their dust-mass injection rate into the ISM of Leo A and Sextans A was determined to be (8.2 $\pm$ 1.8) $\times 10^{-9}$  $M_\odot \,{\rm yr}^{-1}$ and (6.2 $\pm$ 0.2) $\times 10^{-7}$ $M_\odot \,{\rm yr}^{-1}$, respectively. The majority of the dust in each galaxy is produced by a small number of evolved sources. We also find tentative evidence for O-rich dust formation at low metallicity. To confirm the nature of these objects additional high-resolution imaging or spectroscopic data are necessary.

\vspace{0.4cm}

Jones and Meixner acknowledge financial support from NASA grant, NNX14AN06G.
O. Jones was partially supported by an HST grant: STScI 14073.015.  
IM acknowledges support from the UK Science and Technology Facility Council under grant ST/L000768/1.
This research relied on the following resources: NASA's Astrophysics Data System and the SIMBAD and VizieR databases, operated at the Centre de Donn\'{e}es astronomiques de Strasbourg.
This  research made use of  Astropy,  a  community-developed core Python package for Astronomy \citep{Astropy2013}.

Facilities: WIYN:0.9m WHIRC


\input{journaldefs}

\input{ms.bbl}
\appendix

\end{document}

%% file: journaldefs.tex

\def\aj{AJ}					
\def\actaa{Acta Astron.}                        
\def\araa{ARA\&A}				
\def\apj{ApJ}					
\def\apjl{ApJL}					
\def\apjs{ApJS}					
\def\ao{Appl.~Opt.}				
\def\apss{Ap\&SS}				
\def\aap{A\&A}					
\def\aapr{A\&A~Rev.}				
\def\aaps{A\&AS}				
\def\azh{AZh}					
\def\baas{BAAS}					
\def\jrasc{JRASC}				
\def\memras{MmRAS}				
\def\mnras{MNRAS}				
\def\pra{Phys.~Rev.~A}				
\def\prb{Phys.~Rev.~B}				
\def\prc{Phys.~Rev.~C}				
\def\prd{Phys.~Rev.~D}				
\def\pre{Phys.~Rev.~E}				
\def\prl{Phys.~Rev.~Lett.}			
\def\pasp{PASP}					
\def\pasj{PASJ}					
\def\qjras{QJRAS}				
\def\skytel{S\&T}				
\def\solphys{Sol.~Phys.}			
\def\sovast{Soviet~Ast.}			
\def\ssr{Space~Sci.~Rev.}			
\def\zap{ZAp}					
\def\nat{Nature}				
\def\iaucirc{IAU~Circ.}				
\def\aplett{Astrophys.~Lett.}			
\def\apspr{Astrophys.~Space~Phys.~Res.}		
\def\bain{Bull.~Astron.~Inst.~Netherlands}	
\def\fcp{Fund.~Cosmic~Phys.}			
\def\gca{Geochim.~Cosmochim.~Acta}		
\def\grl{Geophys.~Res.~Lett.}			
\def\jcp{J.~Chem.~Phys.}			
\def\jgr{J.~Geophys.~Res.}			
\def\jqsrt{J.~Quant.~Spec.~Radiat.~Transf.}	
\def\memsai{Mem.~Soc.~Astron.~Italiana}		
\def\nphysa{Nucl.~Phys.~A}			
\def\physrep{Phys.~Rep.}			
\def\physscr{Phys.~Scr}				
\def\planss{Planet.~Space~Sci.}			
\def\procspie{Proc.~SPIE}			
\let\astap=\aap
\let\apjlett=\apjl
\let\apjsupp=\apjs
\let\applopt=\ao
